\pdfoutput=1 

\documentclass[a4paper, 11pt, final]{article}

\usepackage{amssymb} 
\usepackage{amsthm} 
\usepackage{amsmath,empheq}
\usepackage{bbm}

\DeclareMathAlphabet{\mathcal}{OMS}{cmsy}{m}{n}
\DeclareMathAlphabet\mathbfcal{OMS}{cmsy}{b}{n}

\DeclareFontFamily{U}{dutchcal}{\skewchar\font=45 }
\DeclareFontShape{U}{dutchcal}{m}{n}{<-> s*[1.0] dutchcal-r}{}
\DeclareFontShape{U}{dutchcal}{b}{n}{<-> s*[1.0] dutchcal-b}{}
\DeclareMathAlphabet{\mathcald}{U}{dutchcal}{m}{n}
\SetMathAlphabet{\mathcald}{bold}{U}{dutchcal}{b}{n}
\DeclareMathAlphabet\mathcalz{T1}{pzc}{mb}{it}

\usepackage[nice]{nicefrac} 
\usepackage{siunitx} 

\usepackage{newtxtext, newtxmath} 

\usepackage{anyfontsize}
\usepackage[utf8]{inputenc}
\usepackage[T1]{fontenc}
\usepackage{microtype} 

\providecommand{\JEL}[1]{\textit{\textbf{JEL: }} #1}
\providecommand{\keywords}[1]{\textbf{\textit{Keywords--- }} #1}

\usepackage{setspace} 

\usepackage{parskip} 
\usepackage{etoolbox}
\usepackage{titlesec}
\titleformat{\section}{\normalfont\Large\bfseries}{\thesection}{1em}{}
\titleformat{\subsection}{\normalfont\large\bfseries}{\thesubsection.}{1em}{}
\titleformat{\subsubsection}{\normalfont\normalsize\itshape}{\thesubsubsection.}{1em}{}

\usepackage{abstract}

\renewenvironment{abstract}
 {\normalfont
  \begin{center}
  \bfseries \abstractname\vspace{-.5em}\vspace{0pt}
  \end{center}
  \list{}{
    \setlength{\leftmargin}{0cm}%
    \setlength{\rightmargin}{\leftmargin}%
  }%
  \item\relax}
 {\endlist}

\usepackage{authblk} 
\usepackage[bottom]{footmisc} 

\usepackage{multicol} 

\usepackage{enumitem} 

\usepackage{graphicx} 
\usepackage{float} 
\usepackage{subcaption}
\usepackage{afterpage} 

\usepackage{printlen}

\usepackage[labelfont=bf]{caption}
\usepackage{color, colortbl}
\definecolor{LightGray}{rgb}{0.93,0.914,0.914}    
\usepackage{soul} 
\usepackage{longtable,rotating} 
\usepackage{booktabs} 
\usepackage{multirow} 
\usepackage{arydshln} 
\usepackage{bigdelim} 

\usepackage[most]{tcolorbox}            

\PassOptionsToPackage{hyphens}{url} 

\usepackage{fancyhdr}
\fancyheadoffset{0cm}
\makeatletter
\newcommand{\quickwordcount}[1]{
  \immediate\write18{texcount -quiet -incbib -sub=none -utf8 -1 -sum -merge -encoding=utf8 #1.tex > #1-words}%
  \immediate\openin\somefile=#1-words
  \read\somefile to \@@localdummy
  \immediate\closein\somefile
  \setcounter{wordcounter}{\@@localdummy}
  \@@localdummy
}
\makeatother

\usepackage{silence}
\usepackage[style=apa,backend=biber,natbib,hyperref]{biblatex}
\DeclareLanguageMapping{british}{british-apa}
\defbibenvironment{bibliography}{\enumerate}{\endenumerate}{\item}

\usepackage[colorlinks=true,allcolors=gray]{hyperref} 
\let\orgautoref\autoref

\renewcommand{\autoref}[1]
{%
\def\equationautorefname{Eq.}%
\def\figureautorefname{Fig.}%
\def\subfigureautorefname{Fig.}%
\orgautoref{#1}%
}

\usepackage[nameinlink,capitalise]{cleveref}

\usepackage{algorithm,algpseudocode}

\makeatletter
\newlength{\trianglerightwidth}
\algnewcommand{\LineCommentCont}[1]{\Statex \hskip\ALG@thistlm%
  \parbox[t]{\dimexpr\linewidth-\ALG@thistlm}
{\leftskip=\algorithmicindent
  \hangindent=\algorithmicindent 
  \hangafter=1%
  \strut\makebox[\algorithmicindent][c]{$\triangleright$}#1\strut}
  } 
\makeatother

\usepackage[left=1.8cm, right=1.8cm, bottom=2.5cm, top=2.5cm]{geometry}

\begin{document}


\renewcommand{\figureautorefname}{Fig.}
\onehalfspacing



\newcommand{\MainTitleText}{Defining and comparing SICR-events for classifying impaired loans under IFRS 9}


\title{\fontsize{20pt}{0pt}\selectfont\textbf{\MainTitleText
}}


\author[,a,b]{\large Arno Botha \thanks{ ORC iD: 0000-0002-1708-0153; Corresponding author: \url{arno.spasie.botha@gmail.com}}}
\author[,a]{\large Esmerelda Oberholzer \thanks{ ORC iD: 0000-0003-1985-7314; \url{essie.oberholzer11@gmail.com}}}
\author[,a,b]{\large Janette Larney \thanks{ ORC iD: 0000-0003-0091-9917; \url{janette.larney@nwu.ac.za}}}
\author[,a,b]{\large Riaan de Jongh \thanks{ORC iD: 0000-0002-7979-256X; \url{riaan.dejongh@nwu.ac.za}}}
\affil[a]{\footnotesize \textit{Centre for Business Mathematics and Informatics, North-West University, Private Bag X6001, Potchefstroom, 2520, South Africa}}
\affil[b]{\footnotesize \textit{National Institute for Theoretical and Computational Sciences (NITheCS), Stellenbosch 7600, South Africa}}
\renewcommand\Authands{, and }

    

\makeatletter
\renewcommand{\@maketitle}{
    \newpage
     \null
     \vskip 1em%
     \begin{center}%
      {\LARGE \@title \par
      	\@author \par}
     \end{center}%
     \par
 } 
 \makeatother
 
 \maketitle

{
    \setlength{\parindent}{0cm}
    \rule{1\columnwidth}{0.4pt}
    \begin{abstract}
    The IFRS 9 accounting standard requires the prediction of credit deterioration in financial instruments, i.e., significant increases in credit risk (SICR).  However, the definition of such a SICR-event is inherently ambiguous, given its current reliance on evaluating the change in the estimated probability of default (PD) against some arbitrary threshold. We examine the shortcomings of this PD-comparison approach and propose an alternative framework for generating SICR-definitions based on three parameters: delinquency, stickiness, and the outcome period. Having varied these framework parameters, we obtain 27 unique SICR-definitions and fit logistic regression models accordingly using rich South African mortgage and macroeconomic data. For each definition and corresponding model, the resulting SICR-rates are analysed at the portfolio-level on their stability over time and their responsiveness to economic downturns. At the account-level, we compare both the accuracy and dynamicity of the SICR-predictions, and discover several interesting trends and trade-offs. These results can help any bank with appropriately setting the three framework parameters in defining SICR-events for prediction purposes. We demonstrate this process by comparing the best-performing SICR-model to the PD-comparison approach, and show the latter's inferiority as an early-warning system. Our work can therefore guide the formulation, modelling, and testing of any SICR-definition, thereby promoting the timeous recognition of credit losses; the main imperative of IFRS 9.
    \end{abstract}
     
     
     \JEL{C31, C44, G21.}
    
    \rule{1\columnwidth}{0.4pt}
}

\noindent Word count (excluding front matter and appendix):  9199 

\noindent Figure count: 13



\newpage

\section{Introduction}
\label{sec:ch1}

It is no easy task to define a SICR-event, or a \textit{significant increase in credit risk} (SICR), which is essentially a binary event. One common approach relies on estimating a loan's default risk, also known as its \textit{probability of default} (PD) where `default' is another type of binary event. Let this PD be denoted by $p_1(x,t)$ given risk information $x$ observed at time $t$ for a specific loan account. A SICR-event can then be defined by comparing $p_1(x,t_r)$ with $p_1(x,t_t)$ between reporting time $t_r$ and initial recognition time $t_t$, which reflects  \S5.5.9 in  IFRS 9 from \citet{ifrs9_2014}. Should this change in risk estimates (or the \textit{magnitude}) exceed some arbitrarily chosen threshold, then a SICR-event is said to have occurred. This approach immediately highlights at least two challenges in establishing whether credit quality has deteriorated significantly. Selecting an appropriate threshold for the magnitude is non-trivial and highly subjective, which is exacerbated by IFRS 9 being principled instead of overly prescriptive. Secondly, any reliance on the point estimate $p_1(x,t)$ tacitly requires a certain degree of accuracy, lest the subsequent comparison become meaningless. However, attaining sufficient accuracy can itself become challenging given the stochastic nature of default risk, especially when considering an ever-changing macroeconomic environment.

We explore an alternative way of identifying SICR-events using predictive modelling instead of a PD-comparison, without diverging from the principles of IFRS 9 (discussed later). In particular, we produce a predictive model (or supervised classifier) that can incorporate both forward-looking and past-due information in predicting a SICR-event. In achieving compliance with \S5.5.9 in  IFRS 9, one of these input variables should then include the change in lifetime PD from the time of original recognition, or at least incorporate the evolution of default risk over time (\S B5.5.12). While our proposed SICR-modelling approach involves no exterior comparison of risk at initial recognition, it is shown later that our approach is insensitive to the PD at initial recognition. Instead, our approach achieves compliance with IFRS 9 since the risk comparison is embedded within the model itself. Our approach also achieves effectiveness since analysis shows that sufficient SICR-related information is found in other input variables.
This modelling approach can therefore balance the prediction task across a multitude of variables, as statistically weighed, instead of relying (solely) on a PD-comparison approach.

However, training any supervised classifier first requires defining the target event, which can itself be challenging. In this regard, we contend that SICR-classification primarily resolves into predicting future delinquency for non-delinquent accounts; i.e., any SICR-event should logically preempt a default-event.
Accordingly, we formulate a concise SICR-framework from which various SICR-definitions can be generated based on pre-default delinquency levels. By varying the framework's three parameters, we obtain a list of viable SICR-definitions. Each resulting definition is then used as the target definition in training a specific classifier from the same input data. Accordingly, SICR-definitions can be implicitly evaluated by comparing the performance of these classifiers against one another. By doing so, we demonstrate the inherent trade-offs amongst the various SICR-definitions themselves. These trade-offs and broad relationships can help banks in selecting a suitable SICR-definition given the unique context of each bank.
Ultimately, our approach relies fundamentally on building a bespoke SICR-model given a particular definition. This model then predicts the risk of future delinquency for non-delinquent loans today, thereby serving as an early-warning system under IFRS 9.

The present study is closest in design to the work of \citet{harris2013a}, \citet{harris2013b}, \citet{botha2021paper1}, and \citet{botha2022paper2}. In particular, \citet{harris2013a} proposed an algorithm (using random forests with data from Barbados) that yields the `best' default definition based on maximising prediction accuracy. When measured in days past due (DPD), these definitions included: 30, 60, and 90 days. This work was later extended in \citet{harris2013b} using Support Vector Machines (SVMs) and included 120 and 150 DPD as additional definitions. In both studies, the author demonstrated that the overall prediction accuracy is significantly affected by the chosen definition of default. In \citet{botha2021paper1} and \citet{botha2022paper2}, a procedure was devised wherein a delinquency threshold is found at which loan recovery (including legal action) is loss-optimal, thereby informing the default definition. Our work differs contextually in that we explore various SICR-definitions (and their underlying parameters) instead of default definitions.

The notion of SICR-events under IFRS 9 is critically reviewed in \autoref{sec:background}, which includes an in-depth examination of the PD-comparison approach and its aforementioned challenges in defining SICR-events. Literature on alternative approaches is then surveyed, followed by examining the support in IFRS 9 for such alternatives. 
In \autoref{sec:method}, we present a simple three-parameter SICR-framework for generating SICR-definitions by sensibly varying its parameters, as illustrated with a few examples. These SICR-definitions are then used in building various supervised classifiers using binary logistic regression; itself reviewed in the appendix. The subsequent modelling results are discussed and compared across SICR-definitions in \autoref{sec:results}, having used residential mortgage data from a large South African bank. We demonstrate various relationships amongst SICR-definitions across a variety of aspects; all of which forms a reusable analytical framework in guiding the selection of a SICR-definition.
Finally, we conclude the study in \autoref{sec:conclusion} with recommendations, and outline avenues of future research. The R-based source code accompanying this study is published on GitHub; see \citet{botha2024sourcecode}.

\section{Towards identifying SICR-events: A critical re-evaluation under IFRS 9}
\label{sec:background}

The recent introduction of IFRS 9 prompted a paradigm shift in the modelling of credit risk. Generally, the value of a financial asset should be comprehensively adjusted over time in line with a bank's (evolving) expectation of credit risk. The principle is to forfeit a portion of income today into a loss provision that ideally offsets amounts that may be written-off tomorrow. Doing so helps to smooth overall earnings volatility, which is itself a central tenet of risk management, as explained in \citet[pp.~38--44]{VanGestel2009book}. IFRS 9 requires that this loss provision be regularly updated based on a statistical model, i.e., the asset's \textit{Expected Credit Loss} (ECL). Given a new ECL-value, a bank adjusts its loss provision either by raising more from earnings or releasing a portion thereof back into the income statement. This ECL-model represents the probability-weighted sum of cash shortfalls that a bank expects to lose over a certain horizon; see \citet[\S 5.5.17--18,~\S B5.5.28--35,~\S B5.5.44--48]{ifrs9_2014}, as well as \citet{xu2016estimating}.

\begin{figure}[ht!]
\centering\includegraphics[width=1\linewidth,height=0.3\textheight]{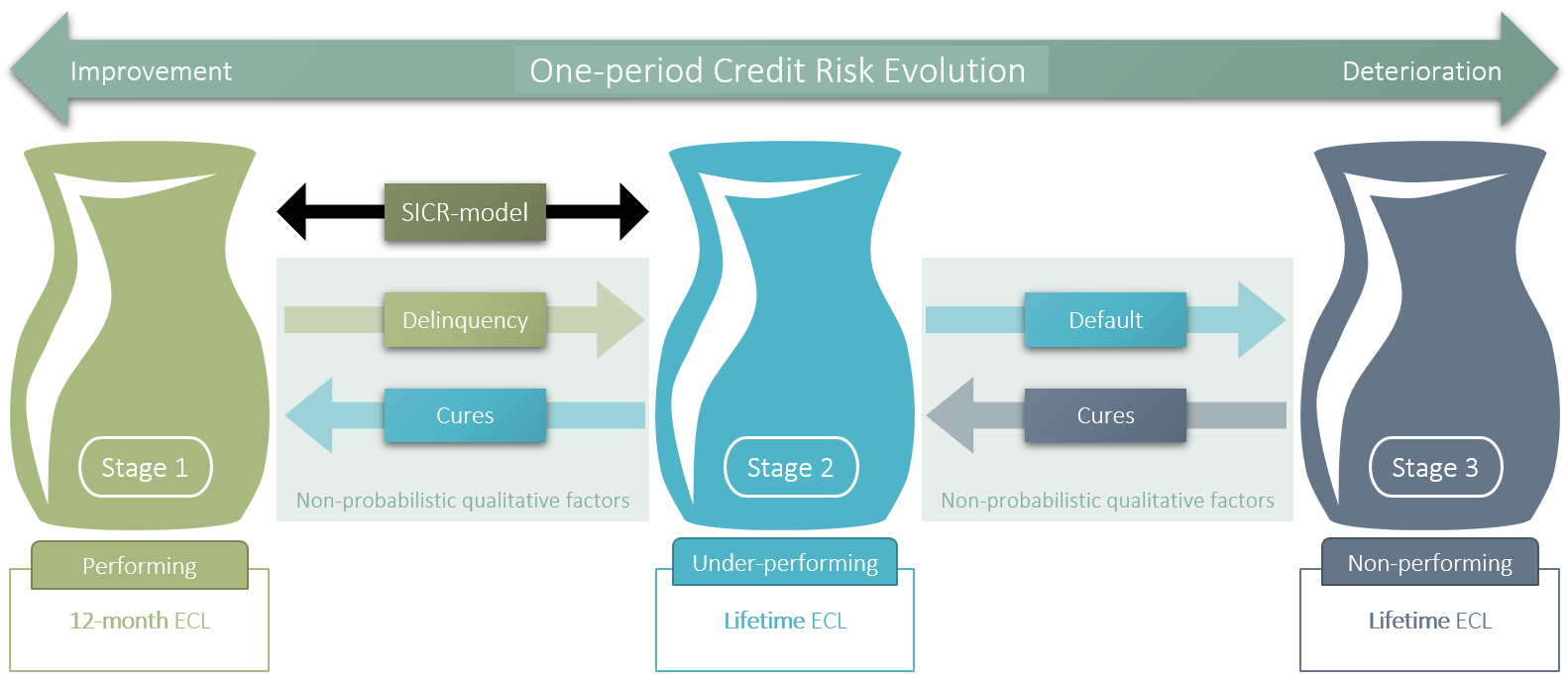}
\caption{Illustrating the one-period evolution of credit risk within the IFRS 9 staged impairment framework. Each subsequent stage implies a greater ECL-estimate to reflect deeper credit deterioration. Arrows indicate possible migrations, subject to meeting certain qualitative criteria. The exception is the probabilistic SICR-component (shaded in dark green), which can include various factors that may predict a SICR-event. From \citet{botha2021phd}.}\label{fig:IFRS9_Stages}
\end{figure}

Regarding the ECL's calculation, IFRS 9 adopts a staged approach in \S 5.5.3 and \S 5.5.5 that is based on the \textit{extent} of the perceived deterioration in the underlying risk. In principle, each of the three stages requires a progressively more severe ECL-estimate, as illustrated in \autoref{fig:IFRS9_Stages}. Stage 1 typically includes most loan assets, provided they either have low credit risk or have not experienced a SICR-event since origination. Stage 2 includes those assets that have deteriorated quite significantly in their credit quality (regardless of measure or SICR-definition), but do not yet qualify as fully credit-impaired (i.e., default); a middle ground of sorts. Lastly, Stage 3 includes those assets with objective evidence of credit impairment, i.e., their future cash flows are likely compromised, e.g., defaulted accounts. These stages can be differentiated from one another by the time horizon of the eventual ECL-estimate: 12 months for Stage 1 and lifetime for Stages 2-3. In particular, a first-stage loss is the portion of lifetime ECL that may occur over the next 12 months, whereas all possible loss-inducing events over the entirety of the asset's remaining life are considered for a second-stage (or third-stage) loss. Together, these stages ought to reflect a more general pattern of deterioration (or improvement) in credit quality over time, which allows for recognising credit losses more timeously; see \S B5.5.2 of IFRS 9, \citet{EY2014}, and \citet{PWC2014}.

Migration between Stage 1 and 2 requires a SICR-component, as conceptualised in \autoref{fig:IFRS9_Stages}. A loan's loss estimate will generally attract a greater provision charge (or coverage rate) when in Stage 2 than in Stage 1. IFRS 9 primarily defines a SICR-event (thus Stage 2) by comparing the PD-estimates $p_1(x,t_r)$ against $p_1(x,t_1)$ at two different points in time $t_r>t_1$, whereupon the magnitude $m(x,t_r)$ is evaluated against a chosen threshold $u > 0$. Note that $m(x,t_r)$ can refer either to the difference $p_1(x,t_r)-p_1(x,t_1)$, or to the ratio $p_1(x,t_r) / p_1(x,t_1)$; though both definitions signify the change in lifetime PD.
If $m(x,t_r)>u$, then the loan is migrated to Stage 2, otherwise it remains in Stage 1. The converse is presumably true as well: a Stage 2 loan is migrated back to Stage 1 once its risk has improved, i.e., if $m(x,t'_r)\leq u$ at some future time $t'_r>t_r$. Doing so would be cost-efficient, particularly since overzealous Stage 2 classification can become prohibitively costly, even if risk-prudent.

However, this \textit{PD-comparison approach} suffers from at least two challenges in identifying SICR-events.
Firstly, the approach presumes that the estimation of PD is indeed accurate; a presumption challenged by \citet{Crook2010} and \citet{chawla2016paper}. In particular, severe model risk is introduced when selecting an inappropriate modelling technique or when failing to capture the time-dynamic nature of lifetime PD. Moreover, the era of big data and associated high-dimensional input spaces are exceptionally challenging when selecting predictive variables; see \citet[\S2.5]{hastie2009elements}. Furthermore, issues concerning data quality (and data preparation) still persist in practice, which means the accuracy of estimation remains questionable. Notwithstanding quality, the paucity of data is another problem when calibrating any technique, perhaps even more so for low default portfolios, as discussed in \citet[\S8]{baesens2016credit}. These issues clearly demonstrate the challenges of producing a single PD-estimate, let alone two.

Secondly, the choice of an appropriate threshold $u$ against which $m(x,t)$ should be evaluated is ambiguous and contentious. Neither IFRS 9 nor most regulators offer any firm guidance on the choice of $u$. While the \citet{EBA_stress2018} defines $u=200\%$, it provides no explanation for this seemingly arbitrary value. In fact, the \citet{pra2019SICR} observed multiple threshold-values that were in use across UK banks and even across different portfolios; all of which attests to further arbitrariness. A single loan portfolio can theoretically even use multiple $u$-values, thereby rendering overall SICR-classification as more risk-sensitive.
The UK-regulator is unsurprised by these differences in SICR-classification, presumably due to the underlying differences across banks in their risk appetites, strategies, and portfolio compositions. It is quite feasible that one bank's SICR-classification will react differently to the same macroeconomic reality, compared to the SICR-classification of a competing bank. 
Given this complexity, the UK-regulator cannot be faulted for expecting greater consistency in the design of SICR-approaches over the longer term, without necessarily ignoring the idiosyncrasies amongst banks or their portfolios.

Some banks select $u$ such that the SICR-flagged population constitutes a pre-defined $x\%$ of the portfolio, where $x\%$ is based on the observed transition rate of becoming delinquent, i.e., reaching 30 days past due. While certainly simple, this method suffers from at least three major drawbacks. 
Firstly, the precise way in which the transition rate is calculated can adversely affect the chosen $u$-value, if done incorrectly. Some notable risk factors include both the length and recency of the underlying sampling window, which may be inappropriately short or exclude known periods of macroeconomic distress; see \citet[\S3.1-2]{botha2021phd} regarding drawbacks of \textit{roll rate analyses}.
Secondly, targeting any $x\%$-value presumes that the delinquent proportion of a portfolio will itself remain largely static in future. This rather crude assumption surely cripples the risk-sensitivity of SICR-classification, especially so during times of macroeconomic upheaval, precisely when SICR-classification should have been dynamic.
Thirdly, stakeholders commonly disagree when adjusting this $x\%$-value across different macroeconomic scenarios, which renders the eventual $u$-value(s) as highly subjective and possibly divorced from reality.

However, IFRS 9 was always intended to be \textit{"principles-based and less complex"} (see \S IN2 in IFRS 9); the lack of firm guidance on the choice of $u$ is therefore unsurprising. Instead of detailed prescriptions, the emphasis is on the purpose behind the rule, which in turn will likely encourage better substantive compliance; see \citet{Black2007}. Accordingly, the lack of prescription regarding SICR-classification seems particularly appropriate, given that it promotes careful evaluation of relevant factors that may influence the individual bank's SICR-classification. Furthermore, the literature is relatively scant regarding the choice of $u$ and is largely limited to corporate lending. 
In particular, \citet{chawla2016paper} introduced three metrics that translate a portfolio's PD term-structure into a measure of spread, which is then used in measuring credit deterioration since origination for SICR-classification. However, not only do these measures depend on observable market prices, but their application still requires appropriate thresholds, with little guidance offered by the authors. 
In contrast, \citet{Ewanchukpaper} proposed that a threshold be found based on the trade-off between income volatility and early default recognition, formulated within a Merton-type framework. That said, the method's success still relies on subjective parameter choices and the availability of market prices. 
Lastly, \citet{brunel2016} suggested an approach for verifying Stage 2 classification by using an underlying PD-model and its accuracy ratio. This approach is centred on maximising the Stage 2 "hit rate", or proportion of SICR-flagged accounts that eventually defaulted. However, this approach is premised on the assumption that all SICR-flagged loans are destined to default. In contrast, a loan's risk profile may very well improve after an initial Stage 2 classification, whereupon it should rightfully cure back to Stage 1. The dynamicity of credit risk and its pre-default evolution over time is therefore completely ignored when simply maximising the Stage 2 "hit rate".

Notwithstanding the previous challenges, \S B5.5.12 in IFRS 9 provides a reprieve. It is not strictly necessary to compare explicit PD-estimates at two points, provided that the evolution of default risk over time is incorporated in some other way. In principle, and when viewed retrospectively, a SICR-event should reasonably preempt a default event such that the timings of both events do not coincide, lest we contravene \S B5.5.21.
This principle suggests using loan delinquency (and its pre-default evolution) directly in defining a SICR-event, at least retrospectively within a dataset. The basis of `SICR-modelling' is then finding a statistical relationship between future SICR-events and a broad set of present-day inputs that predict those SICR-events.
Such a binary classification task can render SICR-predictions more accurately, perhaps using macroeconomic and obligor-specific information, which includes the change in risk since initial recognition; i.e., the magnitude $m(x,t_r)$. In fact, IFRS 9 already requires the use of \textit{"all reasonable and supportable information"} to identify a SICR-event (cf. \S 5.5.4, \S 5.5.9, \S 5.5.11, \S 5.5.17), which further supports statistical modelling.  
Moreover, the PD-comparison approach requires both an accurate PD-model and a suitable threshold $u$, all of which is a relatively indirect way of trying to identify a SICR-event. Instead, SICR-modelling is arguably a more direct approach of classifying future impaired loans into Stage 2, given "\textit{all reasonable and supportable information}" that is observed today and subsequently used as predictive inputs. Therefore, a bespoke SICR-model is likely to be more parsimonious than a PD-model since the inputs of the latter predominantly relate to default risk and not necessarily to the \textit{increase} in credit risk.

Regarding macroeconomic factors, \S B5.5.4 in IFRS 9 already mandates their use in identifying SICR-events. In addition, many authors have found that macroeconomic information can significantly improve PD-prediction; see \citet{Simons2009}, \citet{Bellotti2009}, \citet{bonfim2009}, and \citet{Crook2010}. The work of \citet{leow2016stability} explored default survival models that were trained before and after the 2008 Global Financial Crisis (GFC), which yielded markedly different parametrisations. In turn, the authors explicitly show the dynamic effect (and value) of using macroeconomic information explicitly within PD-prediction. 
A study by \citet{gaffney2019cyclicality} further showed that SICR-classification is highly pro-cyclical and sensitive to economic downturns, at least within the Irish market. The authors followed the PD-comparison approach with $u=200\%$ for SICR-classification, having used Irish residential mortgage data from 2008 to 2015.
These previous studies, together with the IFRS 9 prescription, should bode well for building bespoke SICR-models wherein macroeconomic covariates are explicitly used.

\section{A concise three-parameter SICR-framework for generating SICR-definitions}
\label{sec:method}

A useful starting point for defining a SICR-event is that of a \textit{delinquency measure}, which should quantify the gradual erosion of trust between bank and borrower in honouring the credit agreement. The $g_0$-measure (or the unweighted number of payments in arrears) is selected from \citet{botha2021paper1} for its intuitive appeal and industry-wide ubiquity. 
Now consider an account's $g_0$-measured delinquency over its lifetime $T$, as measured over discrete monthly periods $t=t_1,\dots,T$ from the time of initial recognition $t_1$. In defining a SICR-event, one can compare $g_0(t)$ at time $t$ against a specifiable threshold $d\geq 0$, i.e., $g_0(t)\geq d$. 
In fact, delinquency can be tested over multiple consecutive months, thereby ensuring that a `true' SICR-event is eventually identified at $t$. Such a preliminary SICR-event is said to have occurred at time $t$ if $g_0(v) \geq d$ holds true across a fixed time span $v \in [t-(s-1),t]$. The specifiable parameter $s\geq 1$ is the number of consecutive months for which delinquency is tested; put differently, $s$ is the \textit{stickiness} of the aforementioned delinquency test. These ideas are formalised within the Boolean-valued decision function $\mathcal{G}( d,s,t)$ that yields a binary-valued \textit{SICR-status} in defining a SICR-event at an end-point $t$, expressed as \begin{equation} \label{eq:bool_decision}
 \mathcal{G}(d,s,t) = \left[ \left( \sum_{v=t-(s-1)}^{t}{ \left[g_0(v) \geq d \right]} \right) =s \right] \ \text{for} \ t\geq s \, ,
\end{equation} where $[a]$ are Iverson brackets that outputs 1 if the enclosed statement $a$ is true and 0 otherwise. We illustrate \autoref{eq:bool_decision} for $s=1$ and $s=2$ in \autoref{tab:SICR-decision_illustration} using a hypothetical loan with monthly delinquency observations. For $s=1$, the SICR-status relies on testing $g_0(t)\geq d$ at a single period $t$, which is akin to having no $s$-parameter. For $s=2$, $g_0(t)\geq d$ is tested twice at two consecutive periods $t-1$ and $t$. If both delinquency tests are true, then the resulting sum of the two Iverson statements will equal $s$, thereby signalling a SICR-event at time $t$. The $s$-parameter simply smooths away rapid 0/1-fluctuations in the SICR-status over time, thereby becoming `sticker' as $s$ increases.

\begin{table}[ht!]
\centering
\caption{Illustrating two formulations of the SICR-decision function $\mathcal{G}$ from \autoref{eq:bool_decision} for $(d=1,s=1)$ and $(d=1,s=2)$. Accordingly, SICR-statuses are created using $\mathcal{G}$ for a hypothetical loan and its $g_0$-measured delinquency over time $t$. The $\mathcal{Z}_t(d,s,k)$-process from \autoref{eq:decision_rule_generator} then lags each SICR-status back $k$ periods in creating SICR-outcomes, e.g., $\mathcal{Z}_t(1,1,3)$ at $t=4$ will equate to $\mathcal{G}(1,1,4+3)=1$ three months later in the future.}
\begin{tabular}{p{0.9cm} p{1.7cm} p{1.9cm} p{2.4cm} p{1.9cm} p{2.4cm} p{1.6cm}}
\toprule
Time $t$ & Delinquency $g_0(t)$ & SICR-status $\mathcal{G}(1,1,t)$ & SICR-Outcome $\mathcal{Z}_t(1,1,3)$ & SICR-status $\mathcal{G}(1,2,t)$ & SICR-Outcome $\mathcal{Z}_t(1,2,3)$ & Default $g_0(t)\geq 3$ \\ \midrule
3 & 0 & 0 & 0 &   & 0 & 0 \\
4 & 0 & 0 & 1 & 0 & 0 & 0 \\
5 & 1 & 1 & 1 & 0 & 1 & 0 \\
6 & 0 & 0 & 1 & 0 & 1 & 0 \\
7 & 1 & 1 &   & 0 &   & 0 \\
8 & 2 & 1 &   & 1 &   & 0 \\
9 & 3 & 1 &   & 1 &   & 1\\ \bottomrule
\end{tabular}
\label{tab:SICR-decision_illustration}
\end{table}

\autoref{eq:bool_decision} relies on two specifiable parameters $d$ and $s$ in classifying a loan's accrued delinquency over time $t$. The loan's resulting binary-valued SICR-statuses, i.e., its $\mathcal{G}(d,s,t)$-values, can now be used within a typical cross-sectional modelling setup for predicting future SICR-events, or \textit{SICR-outcomes}. In preparing the modelling dataset, we observe all predictive information of loan $i$ at a particular time $t$. Then, the loan's future SICR-status at time $t+k$ is merged to the observations at $t$, thereby taking a snapshot between two points in time, or a cross-section.
However, the chosen value for this third parameter $k\geq 0$ (or outcome period) can significantly affect modelling results. In particular, both \citet{kennedy2013window} and \citet{mushava2018experimental} examined the outcome period's effect in predicting default risk, using Irish and South African credit data respectively. Too short a horizon yielded overly volatile results, largely due to risk immaturity and/or seasonal effects. Too long a window led to increasingly inaccurate models, in addition to greater asynchronism with market conditions or even the portfolio's risk composition. Since a SICR-event should ideally preempt a default event in reality, our (cross-sectional) study also contends with various parameter choices for $k$. More formally, a process $\mathcal{Z}_t(d,s,k)$ prepares a given loan's monthly performance history by evaluating \autoref{eq:bool_decision} at `future' time $t+k$, though assigns the result to time $t$; see \autoref{tab:SICR-decision_illustration} for an example using $k=3$. Accordingly, and in constituting our SICR-framework, the binary-valued \textit{SICR-outcome} $y_t$ at time $t=t_1,\dots,T-k$ is created from the future SICR-status as \begin{equation} \label{eq:decision_rule_generator}
    \mathcal{Z}_t(d,s,k) \ : \quad y_t = \mathcal{G}(d,s,t+k) \, .
\end{equation}

Various SICR-definitions are generated using the $\mathcal{Z}_t(d,s,k)$-process from \autoref{eq:decision_rule_generator}, simply by systematically varying its parameters $(d,s,k)$. For this study, the parameter space includes: 1) the threshold $d\in\{1,2\}$ of $g_0$-measured delinquency beyond which SICR is triggered; 2) the level of stickiness $s\in\{1,2,3\}$ within the delinquency test; and 3) the choice of outcome period $k\in\left\{3,6,9,12\right\}$ when modelling SICR-outcomes. While the parameter spaces of $d$ and $s$ are appreciatively small, the same luxury does not hold for the outcome period $k$, which can indeed assume many values. Its enumeration is ultimately guided by experimentation and expert judgement in balancing rigour against practicality. That said, more extreme periods of $k>12$ are investigated later in \autoref{sec:outcomePeriods}, though having restricted $d$ and $s$. 
Regardless, the combined parameter space yields 24 different combinations of the triple $(d,s,k)$, as enumerated in \autoref{tab:SICR_Defs}.

\begin{table}[ht!]
\caption{Numbered SICR-definitions, indexed by $j=1,...,24$, and generated by varying the parameters within the $\mathcal{Z}_t(d,s,k)$-process. Definitions are grouped into six classes and shaded accordingly.}
\centering
\begin{subtable}{.48\linewidth}
    \label{tab:SICR_Defs_1}
    \centering
    \begin{tabular}{p{0.25cm} p{1.5cm} p{1.75cm} p{1.5cm} p{1.5cm}}
        \toprule
        \textbf{$j$} & \textbf{Definition}  & \textbf{Delinquency threshold} & \textbf{Stickiness} & \textbf{Outcome period} \\ \midrule
        \rowcolor[HTML]{ECF4FF} 
        1 & 1a(i) & $d \geq 1$ & $s=1$ & $k=3$ \\
        \rowcolor[HTML]{ECF4FF} 
        2 & 1a(ii) & $d \geq 1$ & $s=1$ & $k=6$ \\
        \rowcolor[HTML]{ECF4FF}
        3 & 1a(iii) & $d \geq 1$ & $s=1$ & $k=9$ \\
        \rowcolor[HTML]{ECF4FF} 
        4 & 1a(iv) & $d \geq 1$ & $s=1$ & $k=12$ \\
        \rowcolor[HTML]{E6FFE6} 
        5 & 1b(i) & $d \geq 1$ & $s=2$ & $k=3$ \\
        \rowcolor[HTML]{E6FFE6} 
        6 & 1b(ii) & $d \geq 1$ & $s=2$ & $k=6$ \\
        \rowcolor[HTML]{E6FFE6} 
        7 & 1b(iii) & $d \geq 1$ & $s=2$ & $k=9$ \\
        \rowcolor[HTML]{E6FFE6} 
        8 & 1b(iv) & $d \geq 1$ & $s=2$ & $k=12$ \\
        \rowcolor[HTML]{FFF2E2} 
        9 & 1c(i) & $d \geq 1$ & $s=3$ & $k=3$ \\
        \rowcolor[HTML]{FFF2E2} 
        10 & 1c(ii) & $d \geq 1$ & $s=3$ & $k=6$ \\
        \rowcolor[HTML]{FFF2E2} 
        11 & 1c(iii) & $d \geq 1$ & $s=3$ & $k=9$ \\
        \rowcolor[HTML]{FFF2E2} 
        12 & 1c(iv) & $d \geq 1$ & $s=3$ & $k=12$ \\ \bottomrule
    \end{tabular}
\end{subtable}
\begin{subtable}{.48\linewidth}
    \label{tab:SICR_Defs_2}
    \centering
    \begin{tabular}{p{0.25cm} p{1.5cm} p{1.75cm} p{1.5cm} p{1.5cm}}
        \toprule
        \textbf{\#} & \textbf{Definition}  & \textbf{Delinquency threshold} & \textbf{Stickiness} & \textbf{Outcome period} \\ \midrule
        \rowcolor[HTML]{C0DAFE} 
        13 & 2a(i) & $d \geq 2$ & $s=1$ & $k=3$ \\
        \rowcolor[HTML]{C0DAFE} 
        14 & 2a(ii) & $d \geq 2$ & $s=1$ & $k=6$ \\
        \rowcolor[HTML]{C0DAFE}
        15 & 2a(iii) & $d \geq 2$ & $s=1$ & $k=9$ \\
        \rowcolor[HTML]{C0DAFE} 
        16 & 2a(iv) & $d \geq 2$ & $s=1$ & $k=12$ \\
        \rowcolor[HTML]{B5FFB5} 
        17 & 2b(i) & $d \geq 2$ & $s=2$ & $k=3$ \\
        \rowcolor[HTML]{B5FFB5} 
        18 & 2b(ii) & $d \geq 2$ & $s=2$ & $k=6$ \\
        \rowcolor[HTML]{B5FFB5} 
        19 & 2b(iii) & $d \geq 2$ & $s=2$ & $k=9$ \\
        \rowcolor[HTML]{B5FFB5} 
        20 & 2b(iv) & $d \geq 2$ & $s=2$ & $k=12$ \\
        \rowcolor[HTML]{FFE1BD} 
        21 & 2c(i) & $d \geq 2$ & $s=3$ & $k=3$ \\
        \rowcolor[HTML]{FFE1BD} 
        22 & 2c(ii) & $d \geq 2$ & $s=3$ & $k=6$ \\
        \rowcolor[HTML]{FFE1BD} 
        23 & 2c(iii) & $d \geq 2$ & $s=3$ & $k=9$ \\
        \rowcolor[HTML]{FFE1BD} 
        24 & 2c(iv) & $d \geq 2$ & $s=3$ & $k=12$ \\ \bottomrule
    \end{tabular}
\end{subtable}
\label{tab:SICR_Defs}
\end{table}

Each entry in \autoref{tab:SICR_Defs} can serve as a particular target definition in building a corresponding SICR-model using some technique. Our chosen chosen modelling technique is binary logistic regression, given its ubiquity in credit risk modelling, as reviewed in \autoref{app:logistic}. Each resulting logit-model will therefore yield a probability score for a particular account at each point during its lifetime. 
These $k$-month forward SICR-predictions reasonably approximate their true lifetime counterpart, which can admittedly only be rendered by using more dynamic/complex modelling techniques, e.g., survival analysis. 
However, and as permitted by \S B5.5.13 of IFRS 9, a series of risk estimates from a 12-month PD-model (see \autoref{app:PD_Model}) can approximate a lifetime PD-measure; in which case, we shall similarly appeal to \S B5.5.13 in our work. That said, this appeal might be questionable since the decision to invoke this clause should also be based on the product and its business processes, not only on data patterns or modelling convenience. We shall leave this aspect for future research, simply in the interest of this paper's brevity.


\section{Comparing SICR-definitions using South African mortgage data}
\label{sec:results}

The SICR-modelling results are structured as follows. In \autoref{sec:calibration}, we describe the data and its resampling scheme, the feature selection process, and how the resulting SICR-models are dichotomised towards providing IFRS 9 staging decisions.
In \autoref{sec:outcomePeriods}, we examine the effect of the outcome period $k$ within the 1a-definition class in \autoref{tab:SICR_Defs} (light blue), having included additional outcome periods beyond the usual 12-month boundary. 
Thereafter, the stickiness parameter $s$ is investigated in \autoref{sec:stickiness} for $d=1$ across all $k$ and $s$, i.e., classes 1a-c in \autoref{tab:SICR_Defs} (lighter shades). We further demonstrate in \autoref{sec:delinquency} the futility of using $d=2$ across all $k$ and $s$ in support of the 30 days past due `backstop' of IFRS 9, having analysed the remaining classes 2a-c in \autoref{tab:SICR_Defs} (darker shades).
Lastly, we compare our SICR-modelling approach in \autoref{sec:approach_comparison} to the SICR-decisions given by following the classical PD-comparison approach, which uses the basic PD-model as described in \autoref{app:logistic}.

\subsection{Data calibration: resampling scheme, feature selection, and dichotomisation}
\label{sec:calibration}

SICR-models are trained and validated using a data-rich portfolio of residential mortgages that was provided by a large South African bank. After applying the $\mathcal{Z}_t(d,s,k)$-process from \autoref{eq:decision_rule_generator}, the resulting credit dataset is structured as $\mathcal{D}=\left\{i,t, y_{it}, \boldsymbol{x}_{it} \right\}$ for each SICR-definition in \autoref{tab:SICR_Defs}, having used the same portfolio of $N$ loans. This dataset $\mathcal{D}$ contains month-end observations of each loan $i=1,\dots,N$ over its lifetime $T_i$, i.e., observing $i$ over discrete time $t=t_1,\dots,T_i$ from each account's time of initial recognition $t_1$. Note that the binary-valued SICR-outcome $y_{it}\in\{0,1\}$ indicates at time $t$ whether account $i$ experienced a SICR-outcome exactly $k$ periods later, given a particular SICR-definition. Put differently, $y_{it}$ is created by applying the $\mathcal{Z}_t(d,s,k)$-process on the history of account $i$.
In circumventing computing constraints and for confidentiality purposes, this mortgage portfolio is sub-sampled using two-way stratified sampling across a wide sampling window of January-2007 up to November-2019. Accounts that predate this window's start are retained, together with their subsequent observations throughout this window. As described and tested in \autoref{app:samplingDesign}, this resampling scheme ultimately leads to two non-overlapping datasets for each SICR-definition: a training set $\mathcal{D}_T$ and a validation set $\mathcal{D}_V$. Both sets are partitioned from a subsampled set $\mathcal{D}_S$ that contains about 250,000 observations for each SICR-definition. We deem these datasets as indeed representative of the population at large, which bodes well for deriving SICR-models later that can generalise beyond $\mathcal{D}_T$.

In predicting the SICR-outcome $y_{it}$, consider $\boldsymbol{x}_{it}=\left\{\boldsymbol{x}_i, \boldsymbol{x}_t, \boldsymbol{x}'_{it} \right\}$ within the raw dataset $\mathcal{D}$ as a realised vector of input variables. These variables are thematically grouped as follows: 
\begin{enumerate}
    \item account-level information $\boldsymbol{x}_i$ for loan $i$, e.g., repayment type (debit order, cash);
    \item macroeconomic information $\boldsymbol{x}_t$ at time $t$, e.g., the prevailing inflation rate;
    \item time-dependent behavioural information $\boldsymbol{x}'_{it}$, e.g., the time spent in a performing spell, or the PD-ratio that signifies the change in default risk since initial recognition.
\end{enumerate}
Selecting viable input variables within each SICR-model is mainly achieved by using iterative logistic regressions, often grouped into various mini-themes in distilling insight. This interactive process is guided by experimentation, expert judgement, model parsimony, statistical significance, macroeconomic theory, goodness-of-fit, and prediction accuracy.
Note that in this work, we are ultimately examining the effect of a particular SICR-definition within a broader multi-definition setup. Therefore, and as a last step, the selected features are `standardised' within each definition class in \autoref{tab:SICR_Defs} such that all SICR-models have the same input space per $(d,s)$-tuple across all $k$-values. This `standardisation' should not be confused with rescaling some quantity towards achieving zero mean and unit variance. By standardising the input space, one can therefore ascribe observable patterns in model performance only to variations in the SICR-definition itself, without contending too much with changes in the input space. Highlights of this interactive selection process are given in \autoref{app:inputSpace}, along with the input space (and variable importance) of each SICR-definition. 
Finally, and for each SICR-definition $j=1,\dots,24$, we define the linear predictor $\eta_{jit}$ for observation $it$ as
\begin{equation} \label{eq:SICR_ModelForm}
    \eta_{jit} = \alpha_j + \boldsymbol{\beta}_j^\mathrm{T}\boldsymbol{x}_i + \boldsymbol{\gamma}_j^\mathrm{T}\boldsymbol{x}_t +\boldsymbol{\delta}_j^\mathrm{T}\boldsymbol{x}'_{it} \, ,
\end{equation}
which is then modelled using $g(\mu_{jit})=\eta_{jit}=\log{\left( p_{jit}\left( \boldsymbol{x}_{ji}; \boldsymbol{x}_{t}; \boldsymbol{x}_{jit} \right) / \left(1 - p_{jit}\left( \boldsymbol{x}_{ji}; \boldsymbol{x}_{t}; \boldsymbol{x}_{jit} \right) \right) \right)}$ as the logit link function, where $\mu_{jit}$ denotes the mean SICR-outcome, and $p_{jit}$ represents the conditional SICR-probability. Together with the intercept $\alpha_j$, the estimable parameter vectors $\left\{\boldsymbol{\beta}_j, \boldsymbol{\gamma}_j, \boldsymbol{\delta}_j \right\}$ are found by maximising the likelihood function, as implemented in the R-programming language.


The probabilistic SICR-predictions from logit-models will need to be dichotomised towards rendering impairment staging decisions under IFRS 9, i.e., Stage 1 or 2, which are respectively called a `negative' or `positive' event. An appropriate cut-off $c_{dsk}\in[0,1]$ is therefore required for dichotomising each probability score $p_1^{(j)}(\boldsymbol{x}_{it})\in[0,1]$ from each SICR-model $j$.
Moreover, SICR-outcomes are relatively rare and the consequences of misclassifying positives vs. negatives are intuitively unequal. Under IFRS 9, false negatives $F^-$ should be costlier than false positives $F^+$ in that the former implies the bank has failed to increase its loss provision for those accounts with \textit{increasing} credit risk, i.e., those accounts with an actual future SICR-outcome. Misclassification costs are accordingly assigned as $c_{F^-}=6$ for false negatives and $c_{F^+}=1$ for false positives, which implies an intuitively high cost ratio of $a=6/1$ across all SICR-models. These costs are deduced using expert judgement and experimentation, though can certainly be refined in future work. Given this $a$-value, each $c_{dsk}$-value is then found using the Generalised Youden Index $J_a$, as detailed in \autoref{app:logistic}.
Finally, each SICR-model is dichotomised into the discrete classifier $h$ that yields the class prediction $h(\boldsymbol{x}_{it})=1$ if $p_1(\boldsymbol{x}_{it})>c_{dsk}$ and $h(\boldsymbol{x}_{it})=0$ otherwise.
\subsection{The effect of the outcome period \texorpdfstring{$k$}{Lg} when defining and predicting SICR-events}
\label{sec:outcomePeriods}

In general, SICR-classification should react dynamically to changes in credit risk and its pre-default evolution over time. This dynamicity is even implicit in \S B5.5.2 of IFRS 9, which postulates that a SICR-event should ideally predate an increase in loan delinquency, i.e., the $g_0$-measure. In predicting such events, shorter outcome periods $k$ can demonstrably achieve this dynamicity more readily than longer periods, since the latter is at greater risk of missing short-term fluctuations in $g_0(t)$ between times $t$ and $t+k$. However, the `optimal' choice of this outcome period is yet unclear, as is the very idea of optimality within this SICR-modelling context. 
To help fill this gap, we deliberately vary $k$ from 3 months up to an extreme of 36 months when training our cross-sectional SICR-models, at least within this particular subsection. 
Aside from the $k$-parameter, the two other parameters are kept constant at $d=1$ and $s=1$, which resolves to definition class 1a within \autoref{tab:SICR_Defs}. These two values are relatively benign for the following two reasons. 
Firstly, the underlying SICR-test $g_0(t)\geq d$ from \autoref{eq:bool_decision} suggests that $d=2$ will yield a subset of SICR-outcomes that are already selected by $d=1$. Secondly, $s=1$ implies zero `stickiness' and simplifies the resulting SICR-definition. Both choices of $d$ and $s$ should therefore have minimal interference when studying the effect of $k$, as intended.

\begin{figure}[ht!]
\centering\includegraphics[width=0.78\linewidth,height=0.71\textheight]{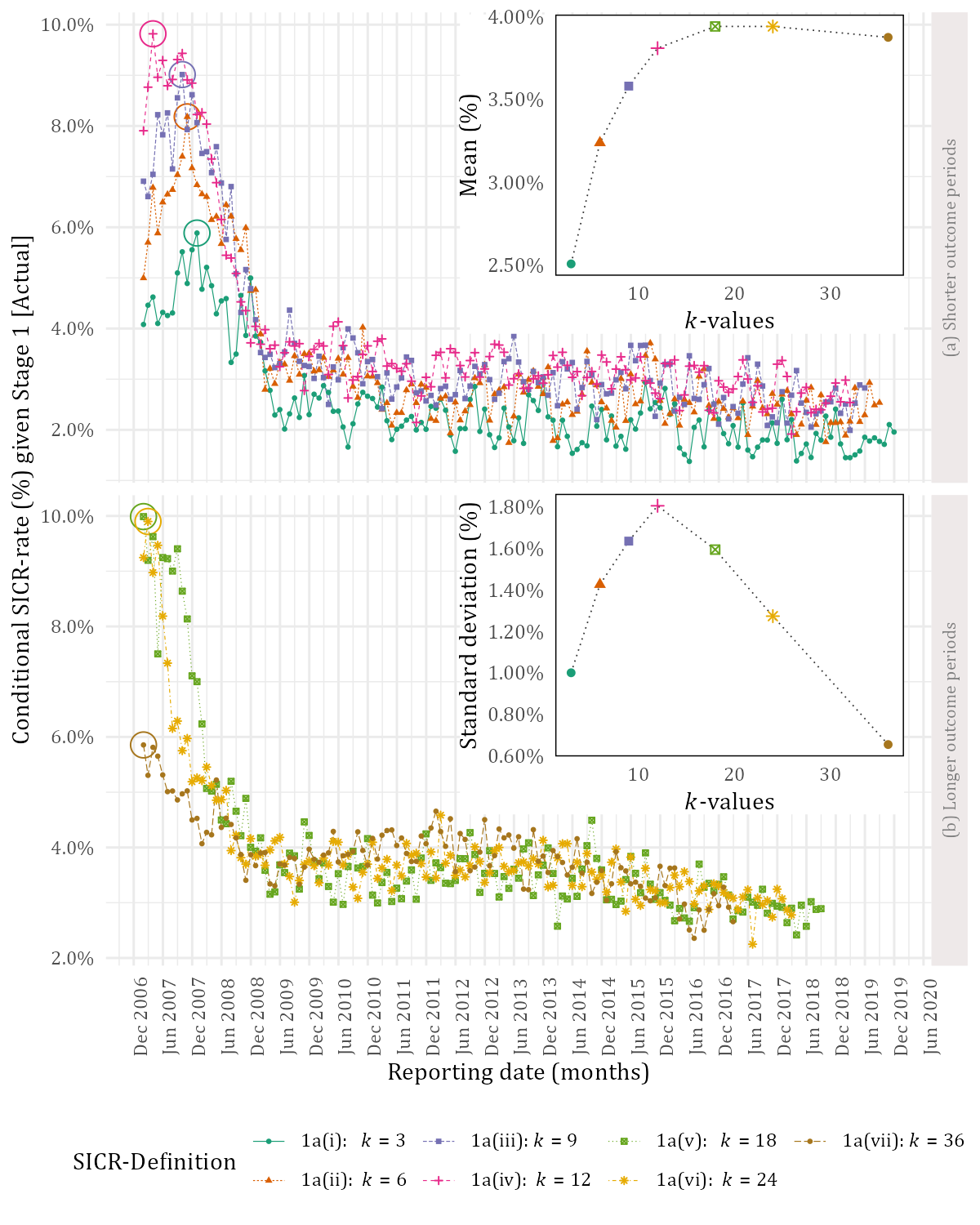}
\caption{Comparing actual SICR-rates over time and across outcome periods $k\in\{3,6,9,12,18,24,36\}$ within $\mathcal{D}_S$ for SICR-definition class 1a from \autoref{tab:SICR_Defs}. The upper panel shows shorter outcome periods while the lower panel shows longer periods, where encircled points denote the maximum. The mean and standard deviation of each resulting time series are summarised within the inset graphs.}\label{fig:1a_SICR_Rates_Actual}
\end{figure}

\begin{figure}[ht!]
\centering\includegraphics[width=0.62\linewidth,height=0.4\textheight]{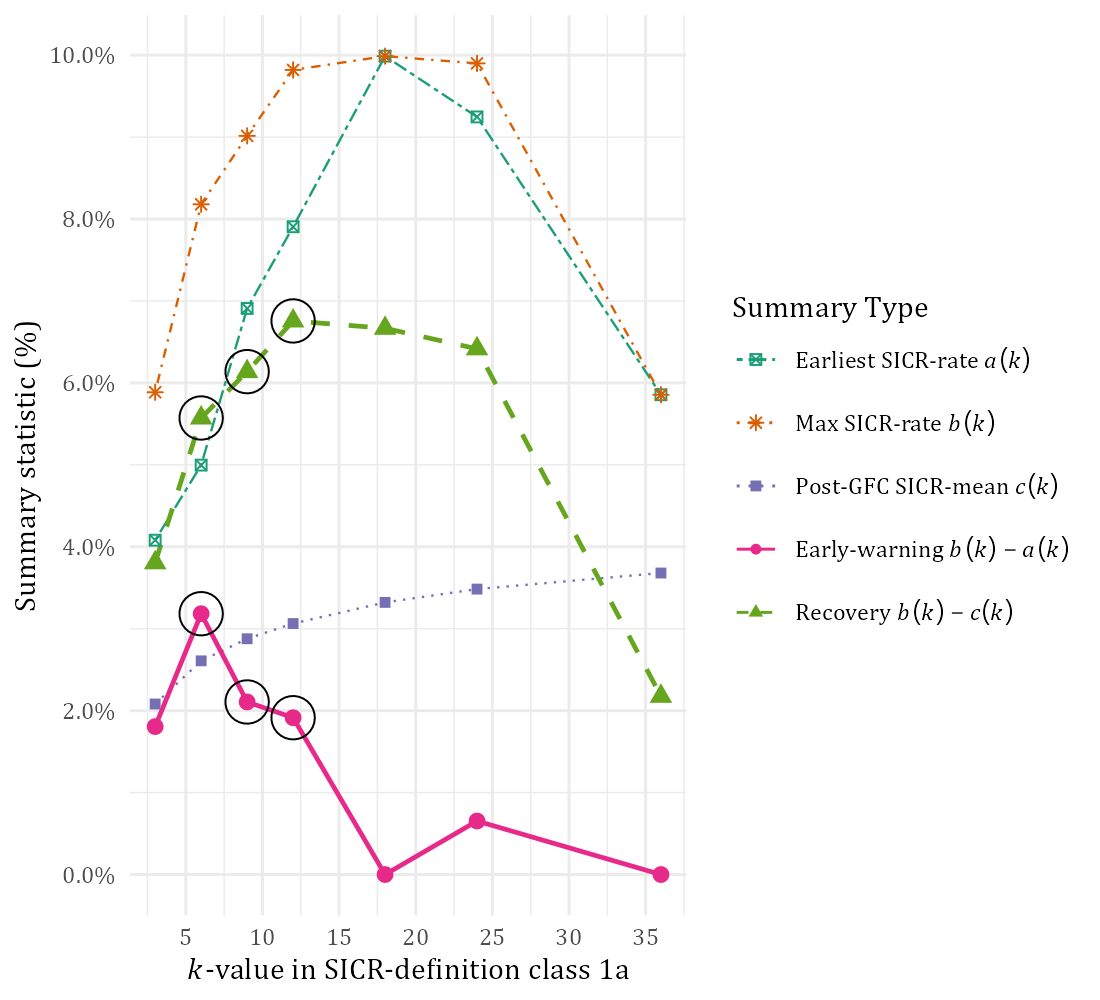}
\caption{Various summary statistics of the actual SICR-rates from \autoref{fig:1a_SICR_Rates_Actual} across chosen outcome periods $k$ for SICR-definition class 1a from \autoref{tab:SICR_Defs}. Summaries include the earliest, maximum, and mean after Dec-2009 (`post-GFC'), as well as differences amongst these summaries, i.e., the early-warning degree and the post-GFC recovery degree. Desirable $k$-values are encircled and discussed.}\label{fig:1a_SICR_Rates_Actual-Summary}
\end{figure}

In exploring the portfolio-level effect of a given SICR-definition, one may start by examining the \textit{actual SICR-rate}; i.e., the Stage 2 transition/delinquency rate over a $k$-month period, as defined in \autoref{app:PerfMeasures}.
In \autoref{fig:1a_SICR_Rates_Actual}, each SICR-rate has a different but increasing mean-level as $k$ increases, especially when examined after the anomalous 2008 Global Financial Crisis (GFC). Since $g_0(t+k)\geq 3 > d$ from \autoref{eq:bool_decision} will hold for both default and SICR-outcomes, larger $k$-values will increasingly capture a greater proportion of defaulting accounts, thereby explaining the phenomenon. Moreover, \autoref{fig:1a_SICR_Rates_Actual} reveals a plateauing effect in the mean, which suggests that choosing $k\geq18$ has a negligible contribution to the overall SICR-mean. At worst, choosing $k\geq 18$ will increasingly select default-instances into the sample, thereby `contaminating' the SICR-mean. Doing so can detract from the very idea of SICR-staging, which should ideally act as a pro-cyclical early warning system for impending credit risk; see \S B.5.5.21 in IFRS 9 from \citet{ifrs9_2014} and \citet{gaffney2019cyclicality}.
The SICR-rate of each $k$-value also exhibits a unique volatility pattern, which is seemingly more stable at the extremes, i.e., $k\leq 3$ and $k > 24$.
However, stable SICR-rates may not necessarily be a useful pursuit, especially not during an unfolding macroeconomic crisis and its subsequent effect on default rates. In particular, the most stable SICR-rates also failed to track the increasing default rates during 2007-2008.
As a working principle for defining the SICR-event, SICR-rates should reasonably exceed default rates since SICR-staging should ideally preempt default. 
This principle avails a useful heuristic in disqualifying both $k\leq 3$ and $k>24$, given their failure in tracking the 12-month default rate; see \autoref{app:PD_Model}.
The actual SICR-rates per $k$-value can be further examined on the basis of various summary statistics, i.e., the earliest $a(k)$, the maximum $b(k)$, and the post-2008 mean $c(k)$. We define two elementary statistics as: 1) the \textit{early warning degree} $b(k)-a(k)$, which denotes the degree to which a SICR-rate can respond to unfolding calamities; and 2) the \textit{recovery degree} $b(k)-c(k)$, which measures the magnitude by which the SICR-rate can normalise post-crisis. Shown across $k$ in \autoref{fig:1a_SICR_Rates_Actual-Summary}, these statistics suggest that $k\in[6,12]$ can yield SICR-rates that reassuringly increase during crises, yet normalise soon thereafter, without becoming overly conservative.

\begin{figure}[ht!]
\centering
\begin{subfigure}{0.49\textwidth}
    \caption{Logistic Regression (Probabilistic)}
    \centering\includegraphics[width=1\linewidth,height=0.34\textheight]{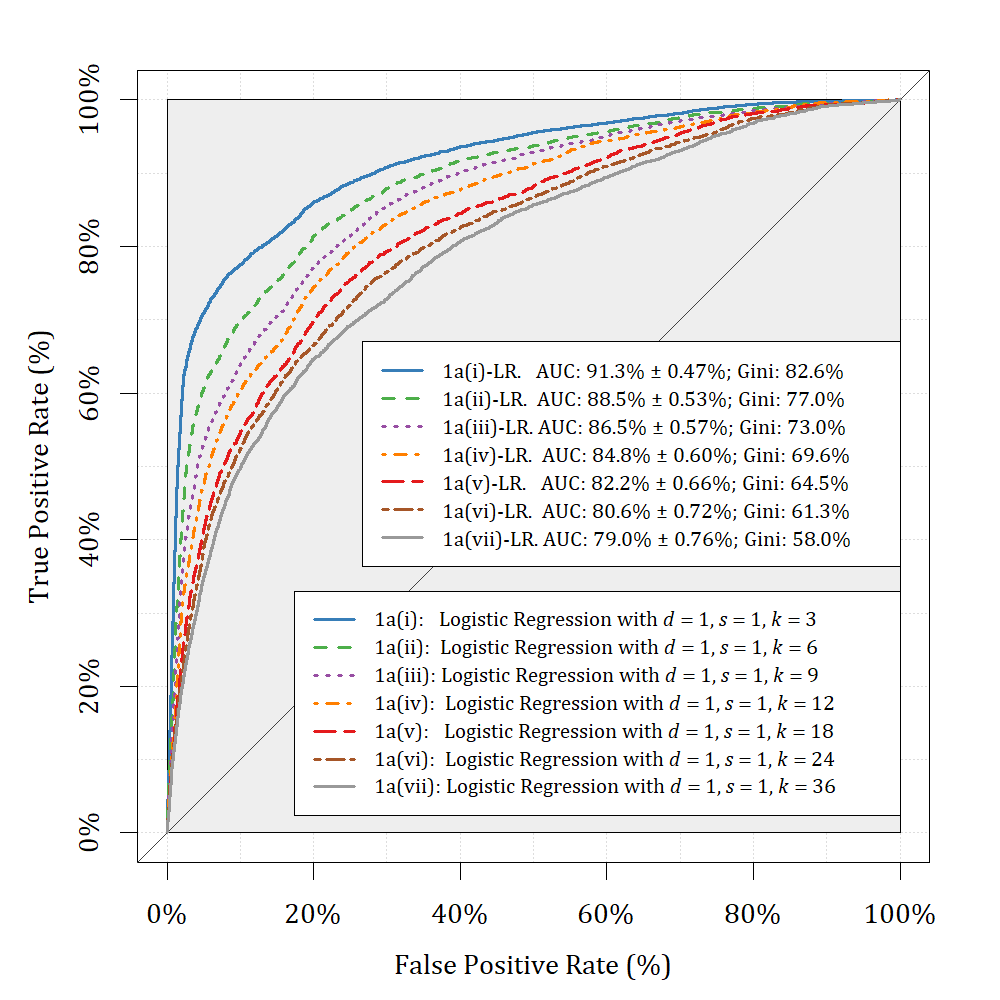}\label{fig:1a-AUC_a}
\end{subfigure}
\begin{subfigure}{0.49\textwidth}
    \caption{Logistic Regression (Discrete)}
    \centering\includegraphics[width=1\linewidth,height=0.34\textheight]{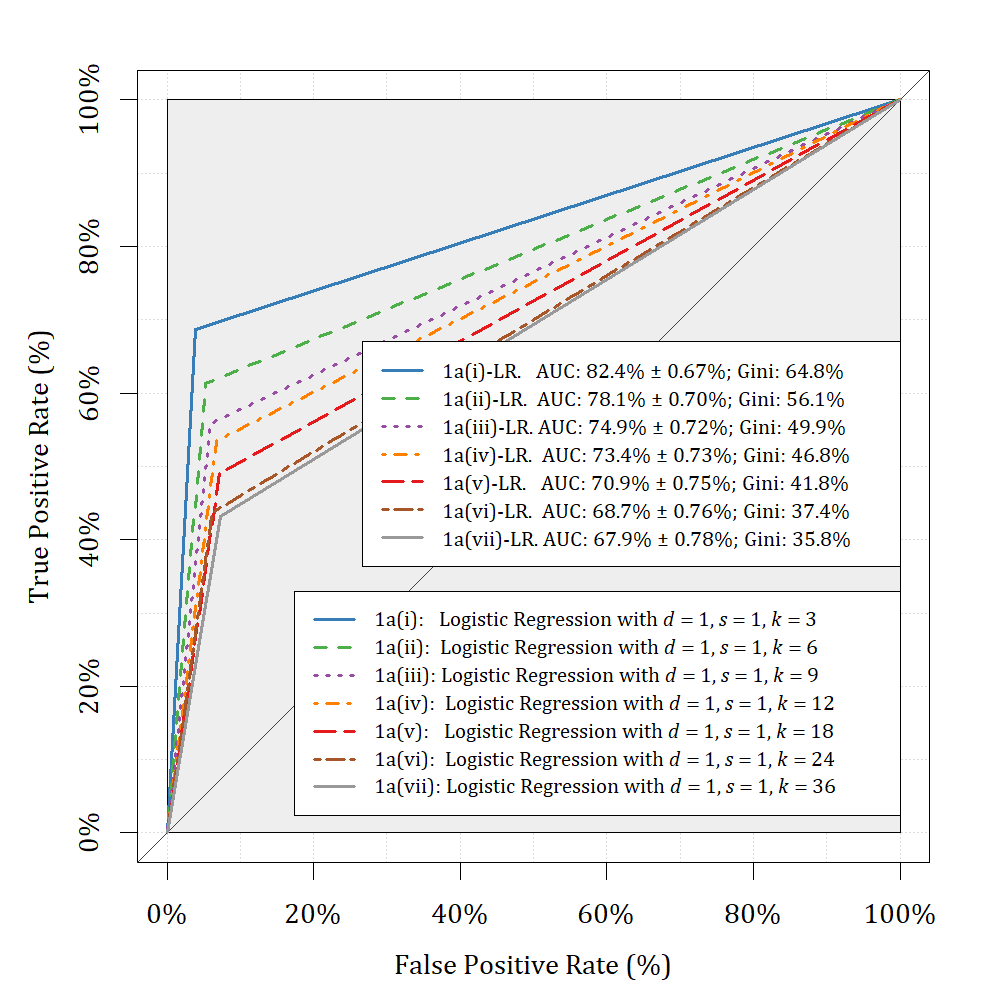}\label{fig:1a-AUC_b}
\end{subfigure}
\caption{ROC-analysis of the SICR-predictions in $\mathcal{D}_V$ resulting from SICR-models developed with definition class 1a ($d=1,s=1$) from \autoref{tab:SICR_Defs}, having varied the outcome period $k$. In \textbf{(a)}, the probability scores from each SICR-model are evaluated as predictions. In \textbf{(b)}, the same scores are dichotomised into discrete predictions using the corresponding cut-off $c_{dsk}$ from \autoref{tab:1a-performance}. The AUC-values are printed, together with 95\% confidence intervals (DeLong-method) and corresponding Gini-values.}\label{fig:1a-AUC}
\end{figure}

In assessing the SICR-model that results from each SICR-definition, we formulate a few performance measures, defined mathematically in \autoref{app:PerfMeasures} though briefly described as follows.
Firstly, the \textit{prevalence} $\phi_{dsk}$ is the proportion of observations that experienced a SICR-event over all time.
Secondly, the \textit{area under the curve} (AUC) summarises a classical ROC-analysis (\textit{receiver operating characteristic}) in measuring a model's discriminatory power, as outlined by \citet{fawcett2006introduction}. 
Thirdly, the \textit{dynamicity} $\omega_{dsk}$, denotes the extent to which SICR-predictions vary over the lifetime of an average account.
Lastly, and as used before, the \textit{instability} $\sigma_{dsk}$ is simply the standard deviation of an actual SICR-rate series.
As summarised in \autoref{tab:1a-performance}, the AUC-values suggest that smaller outcome periods yield more accurate SICR-models than longer periods. This result corroborates the work of \citet{kennedy2013window} and \citet{mushava2018experimental} wherein the outcome period was similarly varied in PD-modelling -- an older `cousin' of SICR-modelling. The plateauing effect in AUC-values suggest yet again that examining smaller $k\leq 18$ values is a more worthwhile endeavour. This trend is mirrored in the discrete AUC-values, having dichotomised the SICR-predictions using $c_{dsk}$ as thresholds; see \autoref{fig:1a-AUC_b}.
Longer outcome periods also result in fewer observed SICR-outcomes, which explains the decreasing $\phi_{dsk}$-values, thereby signifying increased rarity. 
Fewer SICR-outcomes can generally exacerbate the modelling task, which is why AUC decreases as $k$ increases.

\begin{table}[ht!]
\caption{Various performance measures for evaluating SICR-models across different $k$-values within definition class 1a ($d=1,s=1$) from \autoref{tab:SICR_Defs}. AUC-values are given with 95\% confidence intervals (DeLong-method) for both probabilistic and discrete SICR-classifiers, as applied on $\mathcal{D}_V$. The prevalence, dynamicity, and instability measures are calculated within $\mathcal{D}_S$; see \autoref{app:PerfMeasures} for their definitions.}
\label{tab:1a-performance}
\centering
\begin{tabular}{p{1.35cm} p{1.4cm} p{1.5cm} p{2.1cm} p{1.6cm} p{1.4cm} p{1.2cm} p{2.1cm}}
\toprule
\textbf{Definition} & \textbf{Outcome period} $k$ & \textbf{Prevalence} $\phi_{dsk}$ & \textbf{AUC-Probabilistic} & \textbf{Dynamicity} $\omega_{dsk}$ & \textbf{Instability} $\sigma_{dsk}$ & \textbf{Cut-off} $c_{dsk}$ & \textbf{AUC-Discrete}  \\ \midrule
\rowcolor[HTML]{ECF4FF} 
1a(i) & $k=3$ & 6.16\% & 91.3\% \footnotesize{$\pm$ 0.48\%} & 4.3\% & 1.00\% & 12.0\% & 82.4\% \footnotesize{$\pm$ 0.66\%}  \\
\rowcolor[HTML]{ECF4FF} 
1a(ii) & $k=6$ & 6.13\% & 88.5\% \footnotesize{$\pm$ 0.54\%} & 3.8\% & 1.43\% & 10.9\% & 78.1\% \footnotesize{$\pm$ 0.70\%} \\
\rowcolor[HTML]{ECF4FF} 
1a(iii) & $k=9$ & 6.07\% & 86.5\% \footnotesize{$\pm$ 0.57\%} & 3.6\% & 1.64\% & 11.5\% & 74.9\% \footnotesize{$\pm$ 0.73\%} \\
\rowcolor[HTML]{ECF4FF} 
1a(iv) & $k=12$ & 5.99\% & 84.8\% \footnotesize{$\pm$ 0.60\%} & 3.4\% & 1.81\% & 11.6\% & 73.4\% \footnotesize{$\pm$ 0.72\%} \\
\rowcolor[HTML]{ECF4FF} 
1a(v) & $k=18$ & 5.73\% & 82.2\% \footnotesize{$\pm$ 0.67\%} & 2.8\% & 1.60\% & 11.3\% & 70.9\% \footnotesize{$\pm$ 0.74\%} \\
\rowcolor[HTML]{ECF4FF} 
1a(vi) & $k=24$ & 5.46\% & 80.6\% \footnotesize{$\pm$ 0.71\%} & 2.5\% & 1.27\% & 11.3\% & 68.7\% \footnotesize{$\pm$ 0.76\%} \\
\rowcolor[HTML]{ECF4FF} 
1a(vii) & $k=36$ & 5.19\% & 79.0\% \footnotesize{$\pm$ 0.76\%} & 2.0\% & 0.65\% & 9.8\% & 67.9\% \footnotesize{$\pm$ 0.78\%} \\ \bottomrule
\end{tabular}
\end{table}

Smaller outcome periods clearly yield more accurate SICR-models that also produce more dynamic account-level SICR-predictions, as measured by $\omega_{dsk}$. However, this dynamicity may not necessarily translate to the portfolio-level, especially since smaller $k$-values also produce more stable SICR-rates. At the account-level, extremely dynamic SICR-predictions (e.g., $k\leq 3$) can lead to rapid oscillations in moving an account between Stages 1 and 2 over time. This oscillatory effect dampens the overall transition into Stage 2 when aggregating across accounts, hence the less responsive SICR-rate in \autoref{fig:1a_SICR_Rates_Actual-Summary}.
It is therefore questionable to adopt such extremely short outcome periods when, despite their greater prediction accuracy, the associated volatility of the resulting SICR-predictions do not meaningfully translate into more dynamic SICR-rates, as expected at the portfolio-level.
Having disqualified $k\leq 3$, we can similarly disregard $k\geq 18$ given their worsening values in both AUC and $\omega_{dsk}$, and their growing unresponsiveness to externalities such as the 2008-GFC. Together, these results suggest various plateauing effects such that the greatest values of $\omega_{dsk}$ and $\sigma_{dsk}$ occur at $k\in\{6,9,12\}$. The midpoint $k=9$ therefore seems `optimal' when considering the various trade-offs.

\begin{figure}[ht!]
\centering
\begin{subfigure}{0.49\textwidth}
    \caption{1a(i): $k=3$}
    \centering\includegraphics[width=1\linewidth,height=0.28\textheight]{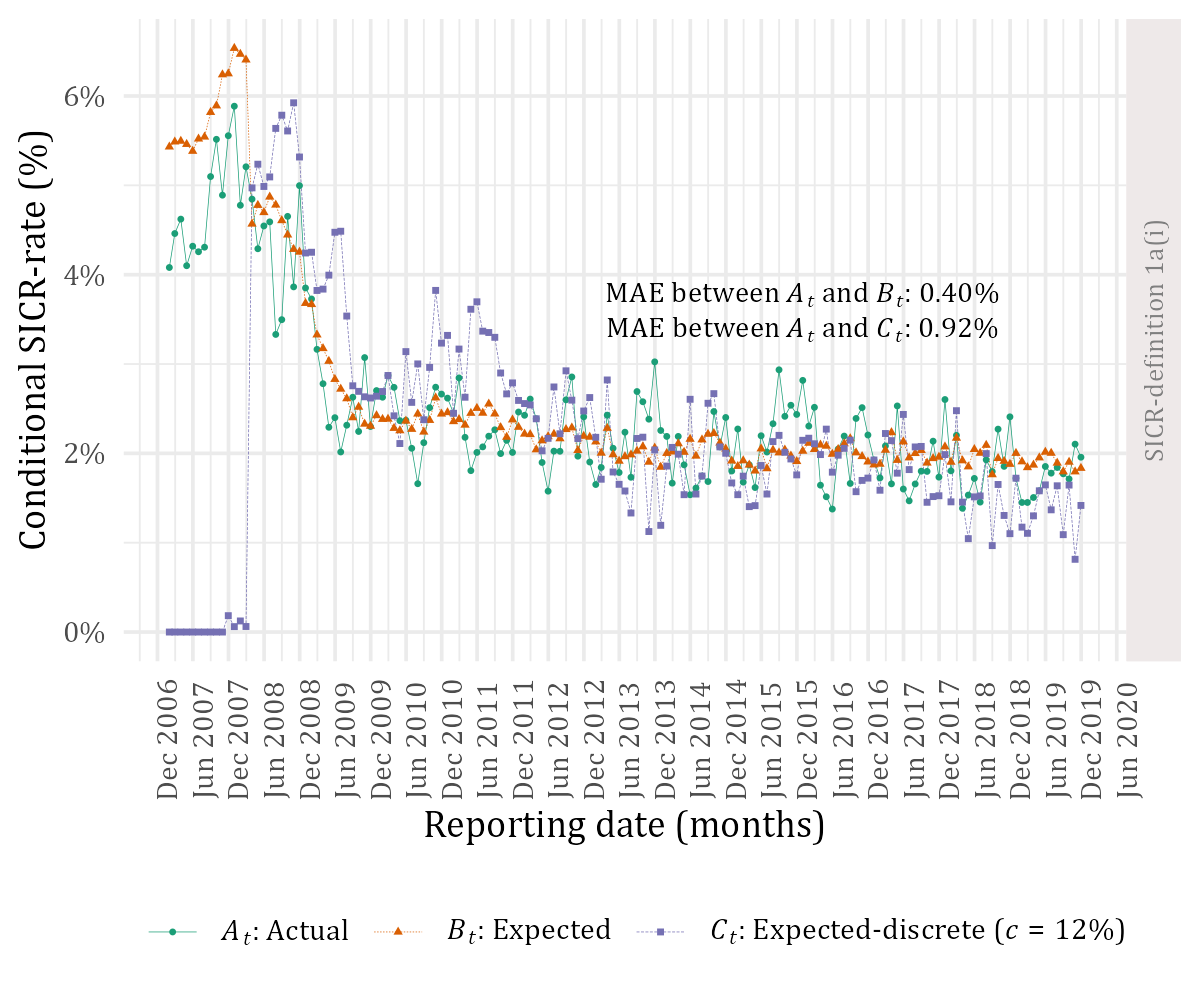}\label{fig:1a_SICR_Rates_a}
\end{subfigure}
\begin{subfigure}{0.49\textwidth}
    \caption{1a(ii): $k=6$}
    \centering\includegraphics[width=1\linewidth,height=0.28\textheight]{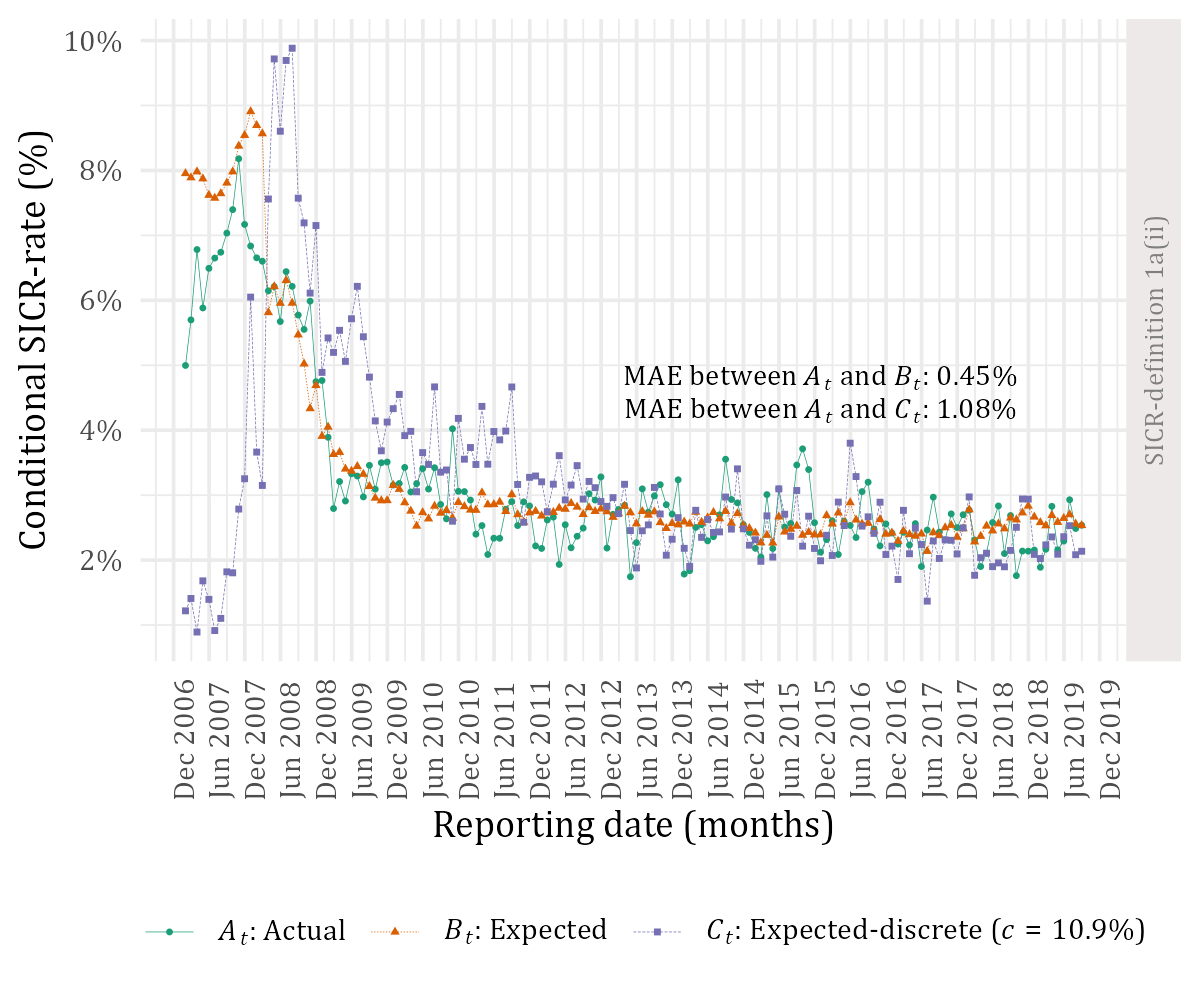}\label{fig:1a_SICR_Rates_b}
\end{subfigure}
\begin{subfigure}{0.49\textwidth}
    \caption{1a(iii): $k=9$}
    \centering\includegraphics[width=1\linewidth,height=0.28\textheight]{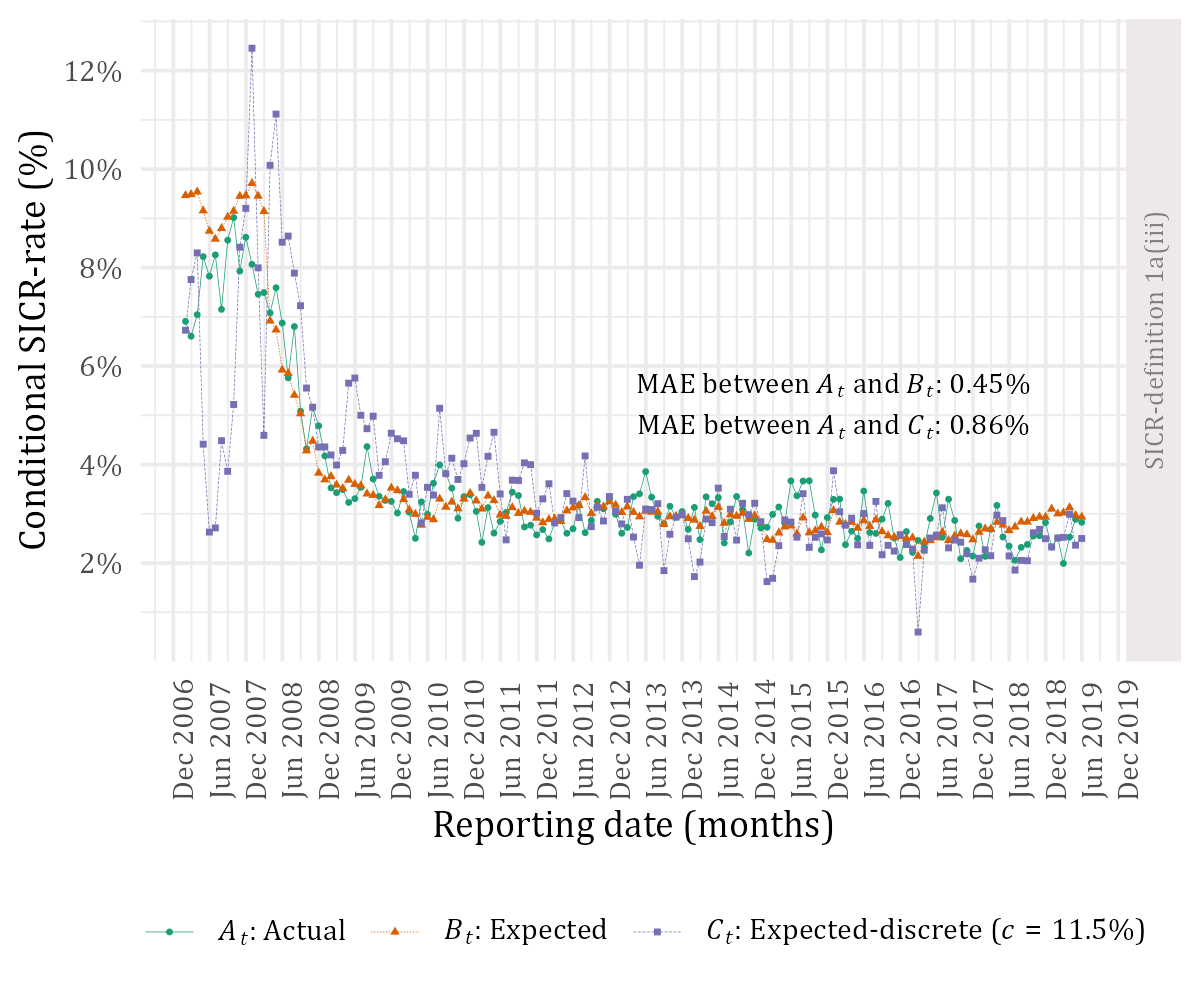}\label{fig:1a_SICR_Rates_c}
\end{subfigure}
\begin{subfigure}{0.49\textwidth}
    \caption{1a(iv): $k=12$}
    \centering\includegraphics[width=1\linewidth,height=0.28\textheight]{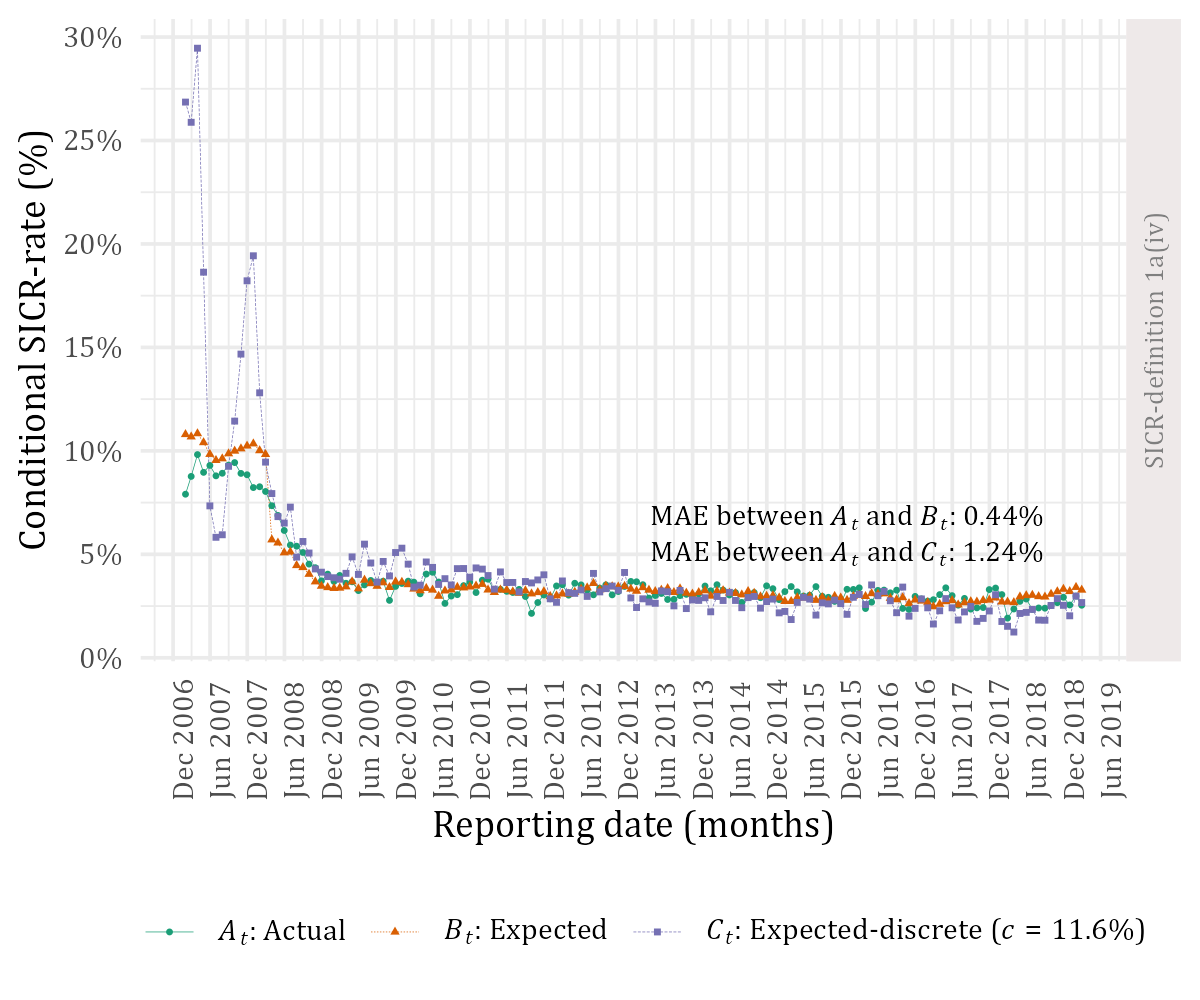}\label{fig:1a_SICR_Rates_d}
\end{subfigure}
\caption{Time graphs of actual $(A_t)$ versus expected $(B_t)$ SICR-rates within $\mathcal{D}_S$ for SICR-definition class 1a ($d=1,s=1$) across shorter outcome periods $k\in \{3,6,9,12\}$; shown in each panel. In creating the discretised expected rate $C_t$, the SICR-model is first dichotomised by using the corresponding cut-off $c_{dsk}$. The MAEs between the actual and each expected SICR-rate are overlaid in summarising the discrepancies over time.}\label{fig:1a_SICR_Rates}
\end{figure}

In \autoref{fig:1a_SICR_Rates}, we show the time graphs of actual $(A_t)$ vs expected $(B_t)$ SICR-rates over calendar time $t$ for $k\in\{3,6,9,12\}$. Ideally, both time graphs should closely overlap each other, which would imply that our aggregated SICR-predictions agree with reality. One can measure the level of such agreement using the \textit{mean absolute error} (MAE) between $A_t$ and $B_t$, denoted as $m_1$ and printed in \autoref{fig:1a_SICR_Rates}. 
Most $m_1$-values are fairly similar with a mean error of 0.44\% across $k$, barring $k\geq 36$. This result corroborates the relatively high AUC-values in \autoref{tab:1a-performance} and visually suggests greater agreement as $k$ increases. 
Having dichotomised the SICR-models, a \textit{discretised} expected SICR-rate $(C_t)$ emerges, which is similarly compared to $A_t$ and summarised again with the MAE $(m_2)$. Clearly, there is more disagreement between either rates, particularly during the 2008-GFC, with a mean $m_2$-value of 1.11\% across $k$; almost three times larger that of $m_1$. Nonetheless, the smallest $m_2$-value still occurred at $k=9$, which further supports its selection as the prudential choice. 
However, these discrete results are highly sensitive to the chosen $c_{dsk}$-value and, by extension, the chosen misclassification cost ratio $a$; all of which can certainly be tweaked in future work.

\subsection{Varying the level of stickiness \texorpdfstring{$s$}{Lg} within SICR-definitions}
\label{sec:stickiness}

\begin{figure}[ht!]
\centering\includegraphics[width=0.78\linewidth,height=0.71\textheight]{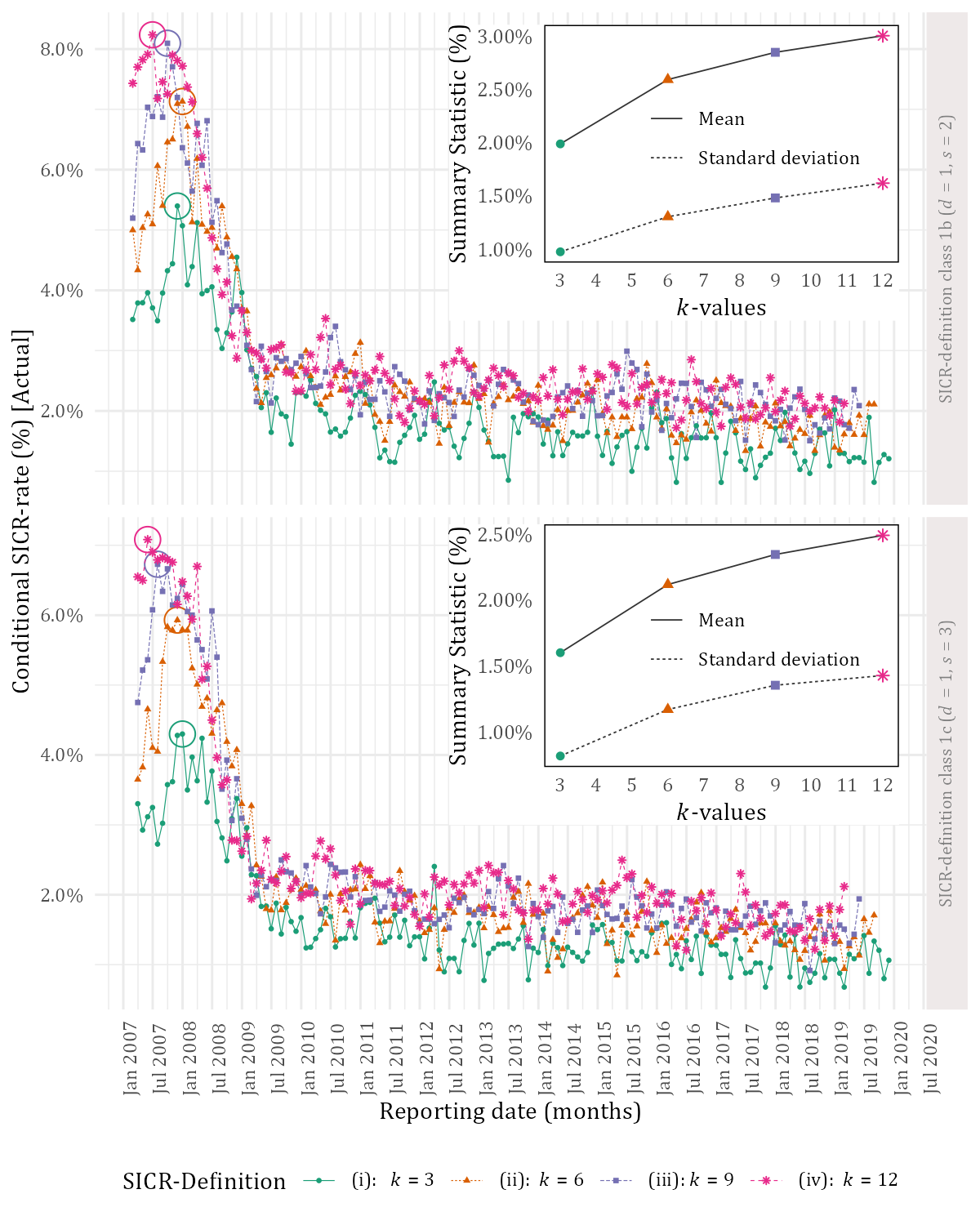}
\caption{Comparing actual SICR-rates over time and across outcome periods $k\in\{3,6,9,12\}$ within $\mathcal{D}_S$ for SICR-definition classes 1b-c from \autoref{tab:SICR_Defs}; one class per panel. Graph design follows that of \autoref{fig:1a_SICR_Rates_Actual}.}\label{fig:1bc_SICR_Rates_Actual}
\end{figure}

In studying the $s$-parameter from \autoref{eq:bool_decision}, recall that it controls the number of account-level delinquency tests that are sequentially conducted over time. The premise of larger $s$-values is to filter out more transient $\mathcal{G}(d,s,t)$-values (or SICR-statuses) that may fluctuate between 0 and 1 over time $t$ for a given account. As $s$ increases, the resulting $\mathcal{Z}_t(d,s,k)$-values (or SICR-outcomes to be predicted) will increasingly equal 1 for more persistent bouts of delinquency. Put differently, overall SICR-classification becomes more deliberate (or less `fickle') for larger $s$-values, which results in SICR-outcomes becoming rarer. This greater scarcity of account-level SICR-outcomes imply that the resulting SICR-rates will decrease on average when aggregating to the portfolio-level, as shown in \autoref{fig:1bc_SICR_Rates_Actual}. 
When matching SICR-rates to those in \autoref{fig:1a_SICR_Rates_Actual} across $k$-values, larger $s$-values clearly deflate both the mean and standard deviation of corresponding SICR-rates. These results attest to the stabilising yet dampening effect of the $s$-parameter in general.
Lastly, and as in \autoref{sec:outcomePeriods}, we calculated both the early-warning and recovery degrees for $s\in\{2,3\}$. The results remained largely similar to that of \autoref{fig:1a_SICR_Rates_Actual-Summary}, though we detected that larger $s$-values seem to inhibit the degree to which SICR-rates can recover from financial crises, likely due to its dampening effect.

The same performance measures from \autoref{tab:1a-performance} are repeated in \autoref{tab:1bc-performance} for evaluating the SICR-models developed using SICR-definition classes 1b-c. Larger $s$-values clearly exacerbate the scarcity caused by larger $k$-values, as evidenced by the progressively lower prevalence $\phi_{dsk}$-rates. We observe the same trends in AUC-estimates across $k$-values, though it appears that stickier SICR-definitions  seemingly produce increasingly more accurate SICR-models. This result suggests that the $s$-parameter succeeds in filtering out more transient SICR-statuses and that larger $s$-values will retain only the more persistent cases of delinquency. In so doing, a clearer more stable statistical relationship can be found amongst input variables, thereby explaining the greater discriminatory power.
However, this greater stability of larger $s$-values naturally (and expectedly) erodes the account-level dynamicity $\omega_{dsk}$; e.g., 3.8\% for $s=1,k=6$ vs 3.2\% for $s=2,k=6$ vs 2.9\% for $s=3,k=6$. Moreover, and when increasing $s$, the great prediction power attained at $k=3$ still fails to coalesce into a similarly dynamic SICR-rate at the portfolio-level, as evidenced by trends in $\sigma_{dsk}$ and \autoref{fig:1bc_SICR_Rates_Actual}.

\begin{table}[ht!]
\caption{Various performance measures for evaluating SICR-models across different $k$-values within definition classes 1b ($d=1,s=2$) and 1c ($d=1,s=3$) from \autoref{tab:SICR_Defs}. Table design follows that of \autoref{tab:1a-performance}.}
\label{tab:1bc-performance}
\centering
\begin{tabular}{p{1.35cm} p{1.4cm} p{1.5cm} p{2.1cm} p{1.6cm} p{1.4cm} p{1.2cm} p{2.1cm}}
\toprule
\textbf{Definition} & \textbf{Outcome period} $k$ & \textbf{Prevalence} $\phi_{dsk}$ & \textbf{AUC-Probabilistic} & \textbf{Dynamicity} $\omega_{dsk}$ & \textbf{Instability} $\sigma_{dsk}$ & \textbf{Cut-off} $c_{dsk}$ & \textbf{AUC-Discrete}  \\ \midrule
\rowcolor[HTML]{E6FFE6} 
1b(i) & $k=3$ & 4.74\% & 93.8\% \footnotesize{$\pm$ 0.43\%} & 3.7\% & 1.07\% & 10.8\% & 85.4\% \footnotesize{$\pm$ 0.72\%}  \\
\rowcolor[HTML]{E6FFE6} 
1b(ii) & $k=6$ & 4.72\% & 89.4\% \footnotesize{$\pm$ 0.56\%} & 3.2\% & 1.38\% & 11.7\% & 78.4\% \footnotesize{$\pm$ 0.80\%} \\
\rowcolor[HTML]{E6FFE6} 
1b(iii) & $k=9$ & 4.68\% & 88.0\% \footnotesize{$\pm$ 0.59\%} & 3.0\% & 1.54\% & 8.6\% & 76.8\% \footnotesize{$\pm$ 0.81\%} \\
\rowcolor[HTML]{E6FFE6} 
1b(iv) & $k=12$ & 4.61\% & 86.5\% \footnotesize{$\pm$ 0.63\%} & 2.8\% & 1.71\% & 10.3\% & 73.8\% \footnotesize{$\pm$ 0.82\%} \\
\rowcolor[HTML]{FFF2E2} 
1c(i) & $k=3$ & 3.82\% & 95.7\% \footnotesize{$\pm$ 0.38\%} & 3.3\% & 0.97\% & 9.7\% & 89.9\% \footnotesize{$\pm$ 0.68\%} \\
\rowcolor[HTML]{FFF2E2} 
1c(ii) & $k=6$ & 3.81\% & 91.6\% \footnotesize{$\pm$ 0.57\%} & 2.9\% & 1.28\% & 9.5\% & 81.7\% \footnotesize{$\pm$ 0.85\%} \\
\rowcolor[HTML]{FFF2E2} 
1c(iii) & $k=9$ & 3.78\% & 88.9\% \footnotesize{$\pm$ 0.64\%} & 2.6\% & 1.44\% & 10.8\% & 76.0\% \footnotesize{$\pm$ 0.91\%} \\ 
\rowcolor[HTML]{FFF2E2} 
1c(iv) & $k=12$ & 3.73\% & 86.7\% \footnotesize{$\pm$ 0.7\%} & 2.4\% & 1.57\% & 8.6\% & 74.0\% \footnotesize{$\pm$ 0.92\%} \\ \bottomrule
\end{tabular}
\end{table}

\begin{figure}[ht!]
\centering
\begin{subfigure}{0.49\textwidth}
    \caption{1b(i): $k=3$}
    \centering\includegraphics[width=1\linewidth,height=0.28\textheight]{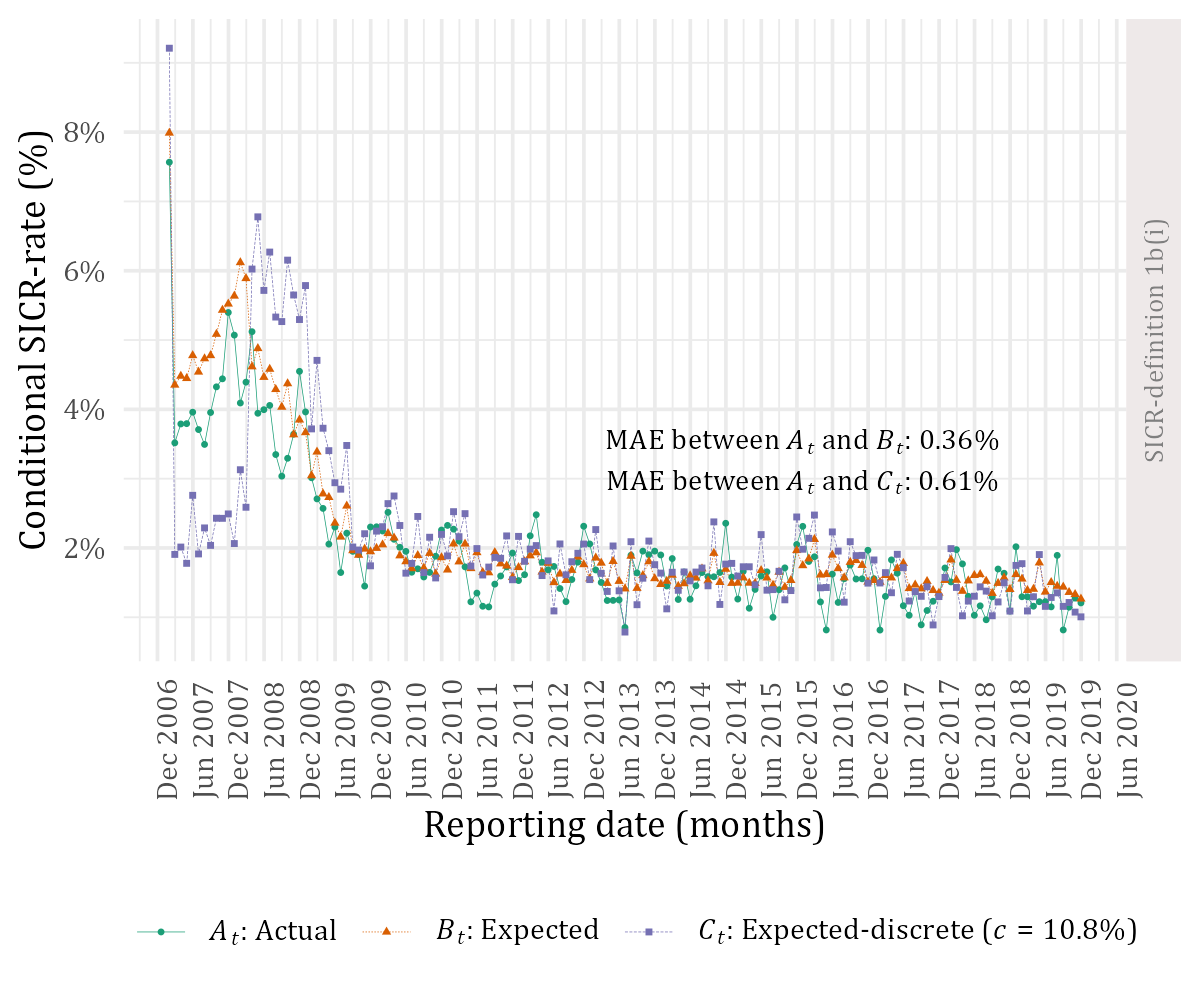}\label{fig:1b_SICR_Rates_a}
\end{subfigure}
\begin{subfigure}{0.49\textwidth}
    \caption{1b(ii): $k=6$}
    \centering\includegraphics[width=1\linewidth,height=0.28\textheight]{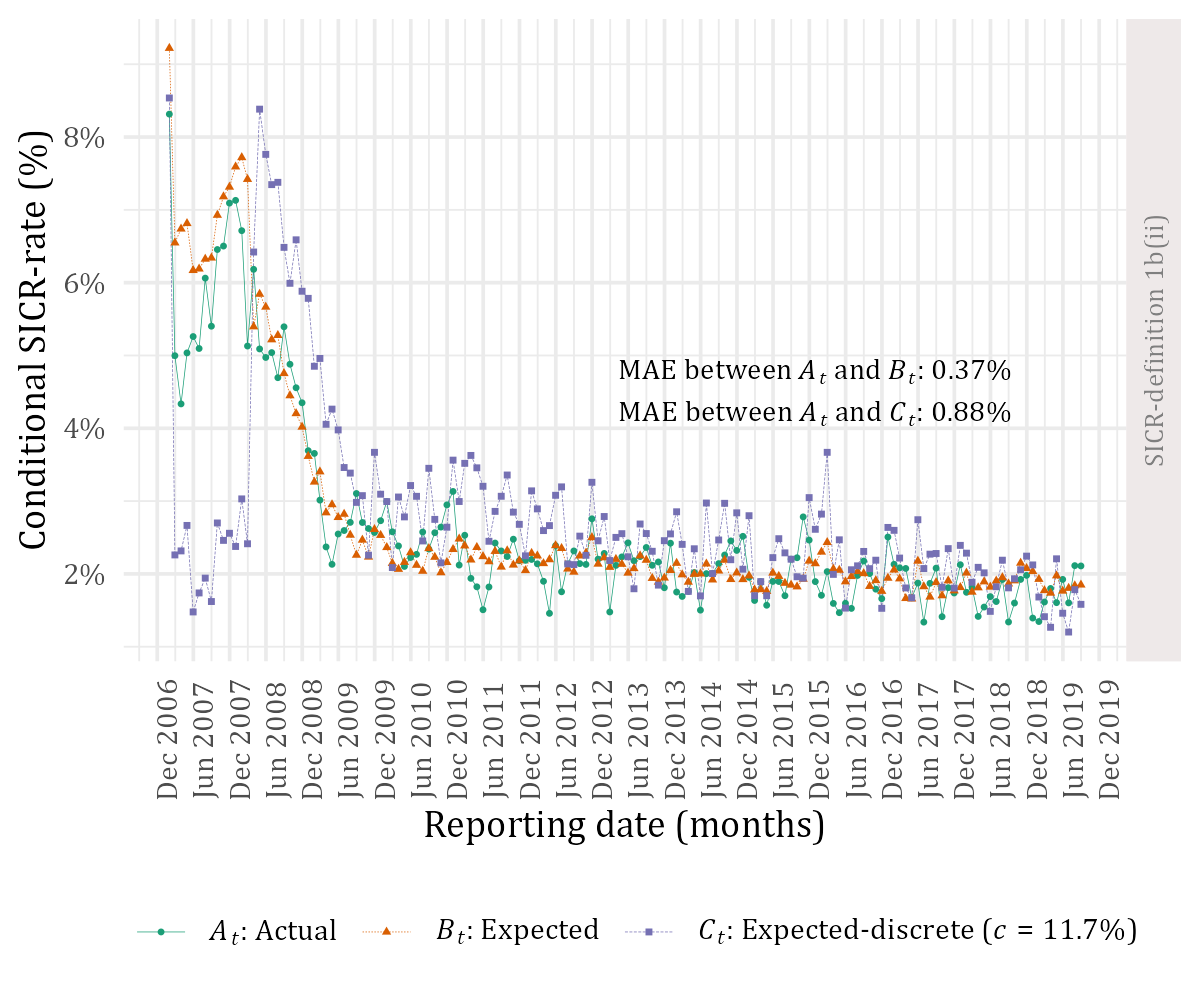}\label{fig:1b_SICR_Rates_b}
\end{subfigure}
\begin{subfigure}{0.49\textwidth}
    \caption{1b(iii): $k=9$}
    \centering\includegraphics[width=1\linewidth,height=0.28\textheight]{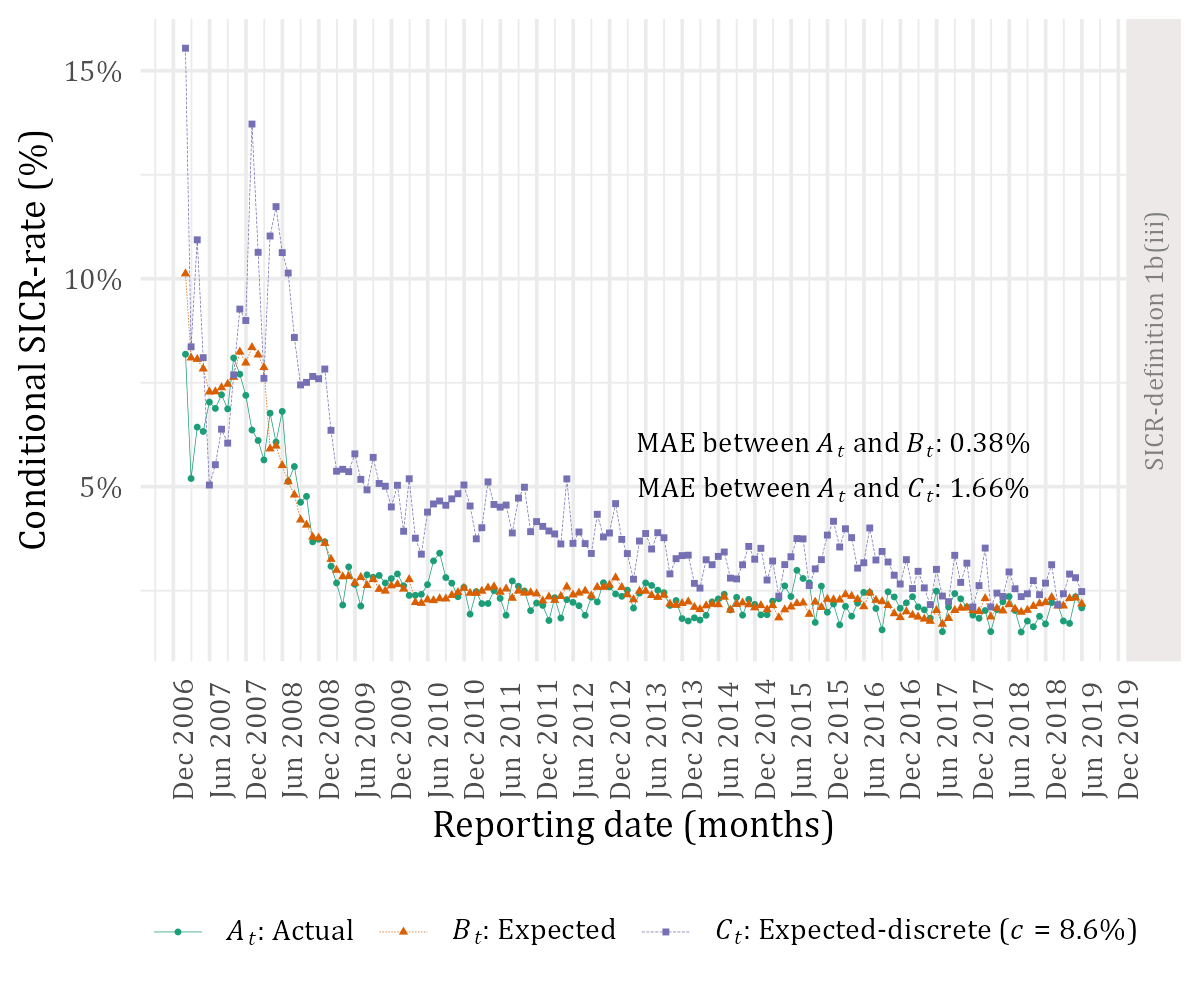}\label{fig:1b_SICR_Rates_c}
\end{subfigure}
\begin{subfigure}{0.49\textwidth}
    \caption{1b(iv): $k=12$}
    \centering\includegraphics[width=1\linewidth,height=0.28\textheight]{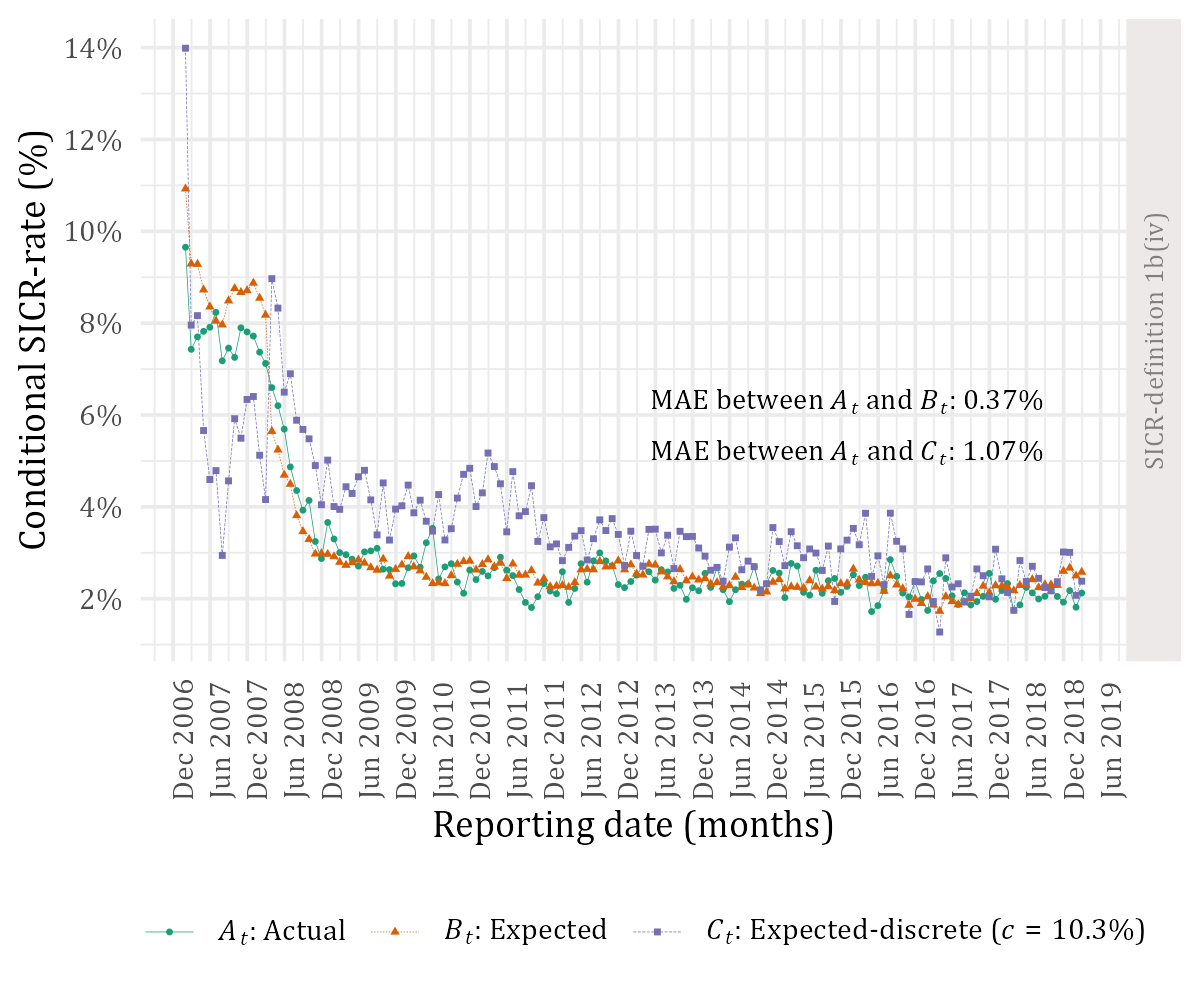}\label{fig:1b_SICR_Rates_d}
\end{subfigure}
\caption{Time graphs of actual versus expected SICR-rates within $\mathcal{D}_S$ for SICR-definition class 1b ($d=1,s=2$) across outcome periods $k\in [3,6,9,12]$. Graph design follows that of \autoref{fig:1a_SICR_Rates}.}\label{fig:1b_SICR_Rates}
\end{figure}

\begin{figure}[ht!]
\centering
\begin{subfigure}{0.49\textwidth}
    \caption{1c(i): $k=3$}
    \centering\includegraphics[width=1\linewidth,height=0.28\textheight]{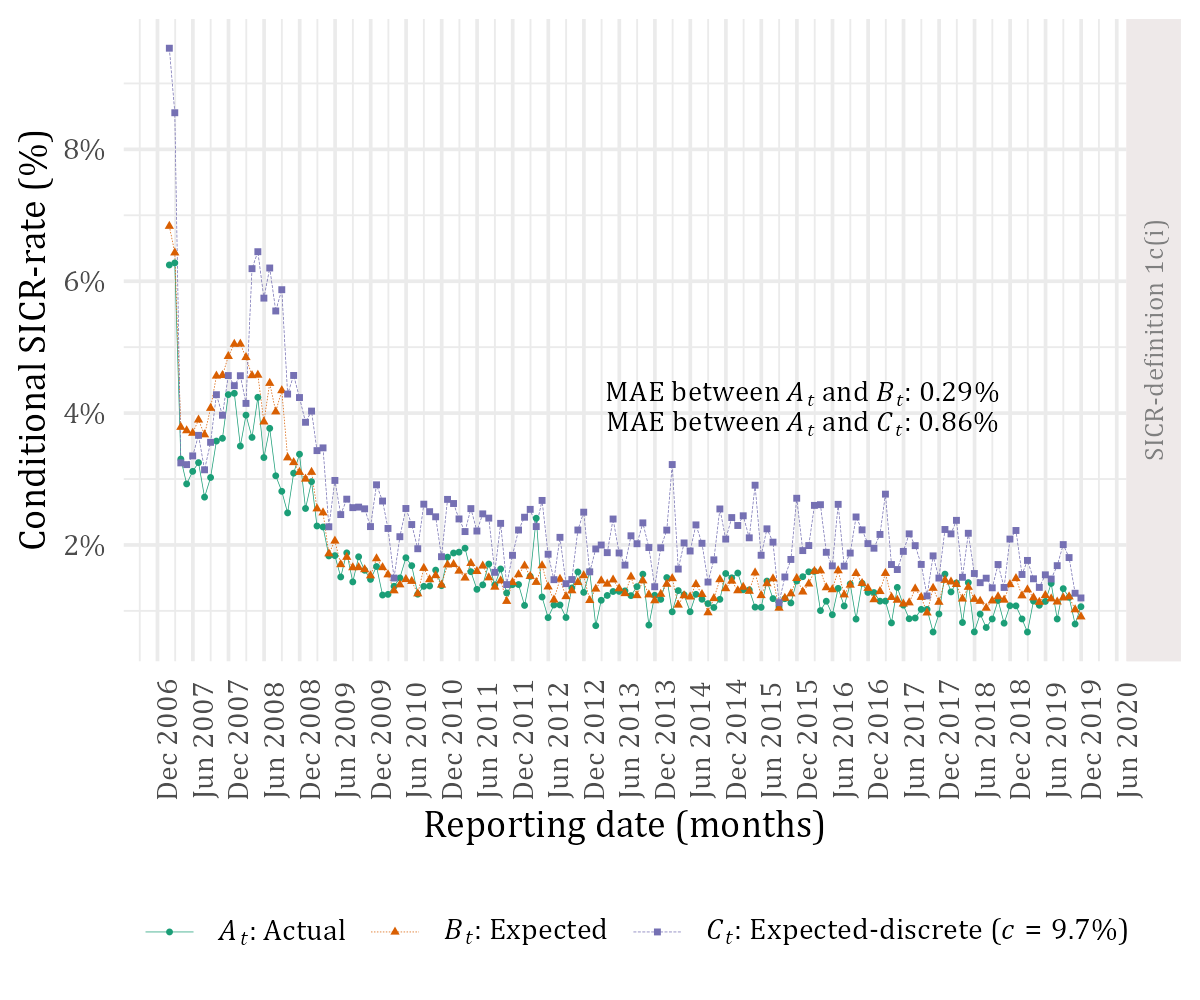}\label{fig:1c-SICR-Inc_a}
\end{subfigure}
\begin{subfigure}{0.49\textwidth}
    \caption{1c(ii): $k=6$}
    \centering\includegraphics[width=1\linewidth,height=0.28\textheight]{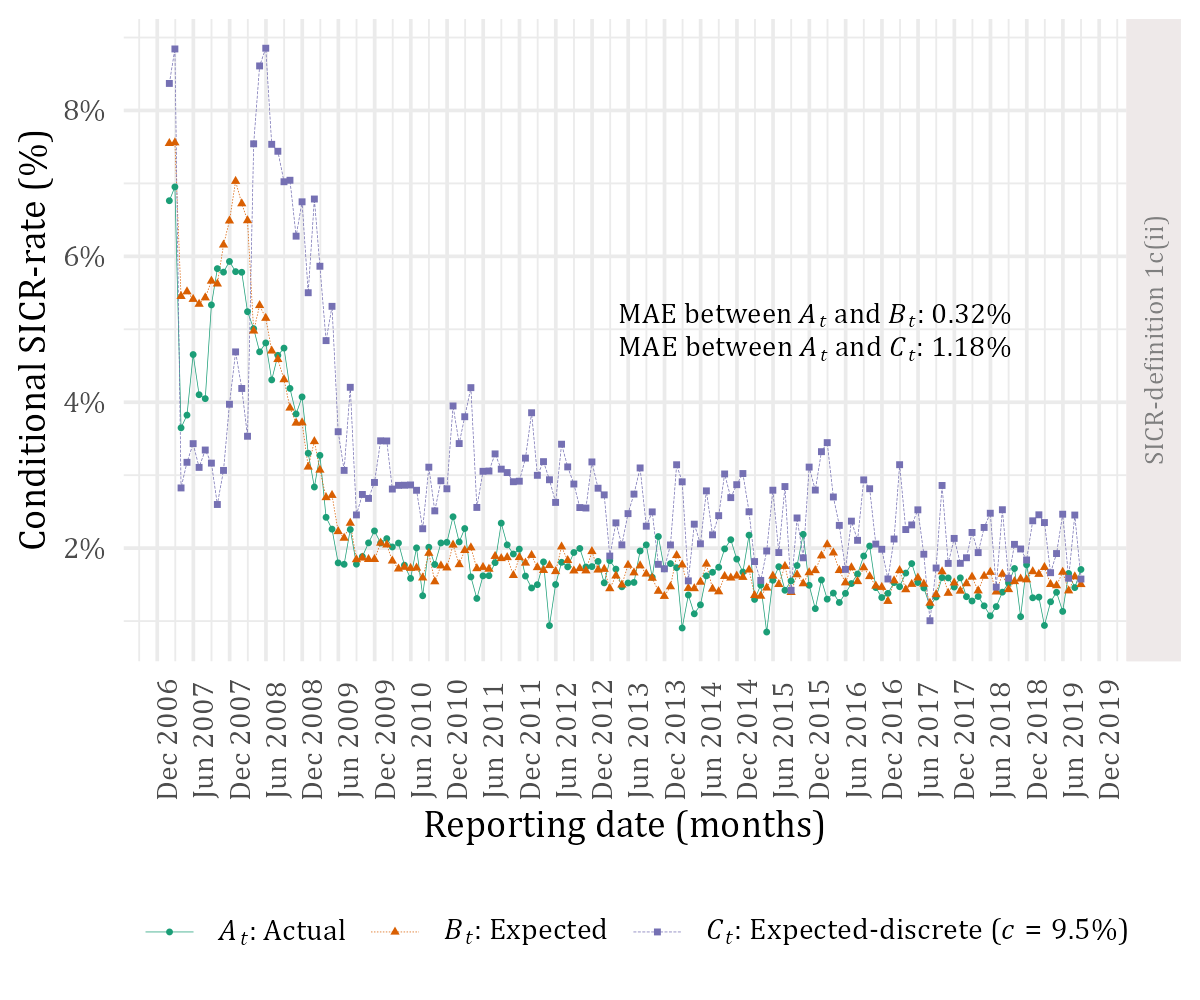}\label{fig:1c-SICR-Inc_b}
\end{subfigure}
\begin{subfigure}{0.49\textwidth}
    \caption{1c(iii): $k=9$}
    \centering\includegraphics[width=1\linewidth,height=0.28\textheight]{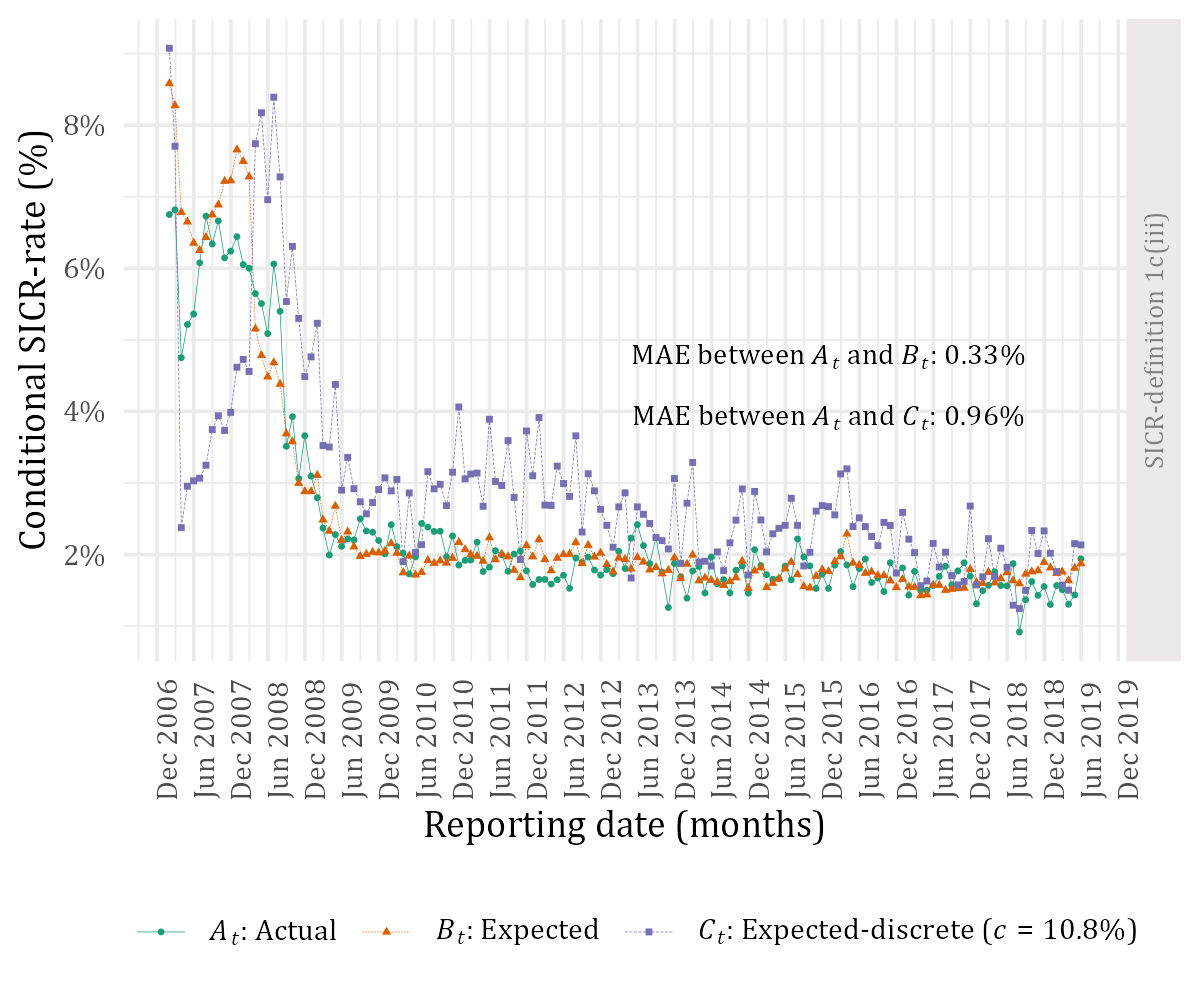}\label{fig:1c-SICR-Inc_c}
\end{subfigure}
\begin{subfigure}{0.49\textwidth}
    \caption{1c(iv): $k=12$}
    \centering\includegraphics[width=1\linewidth,height=0.28\textheight]{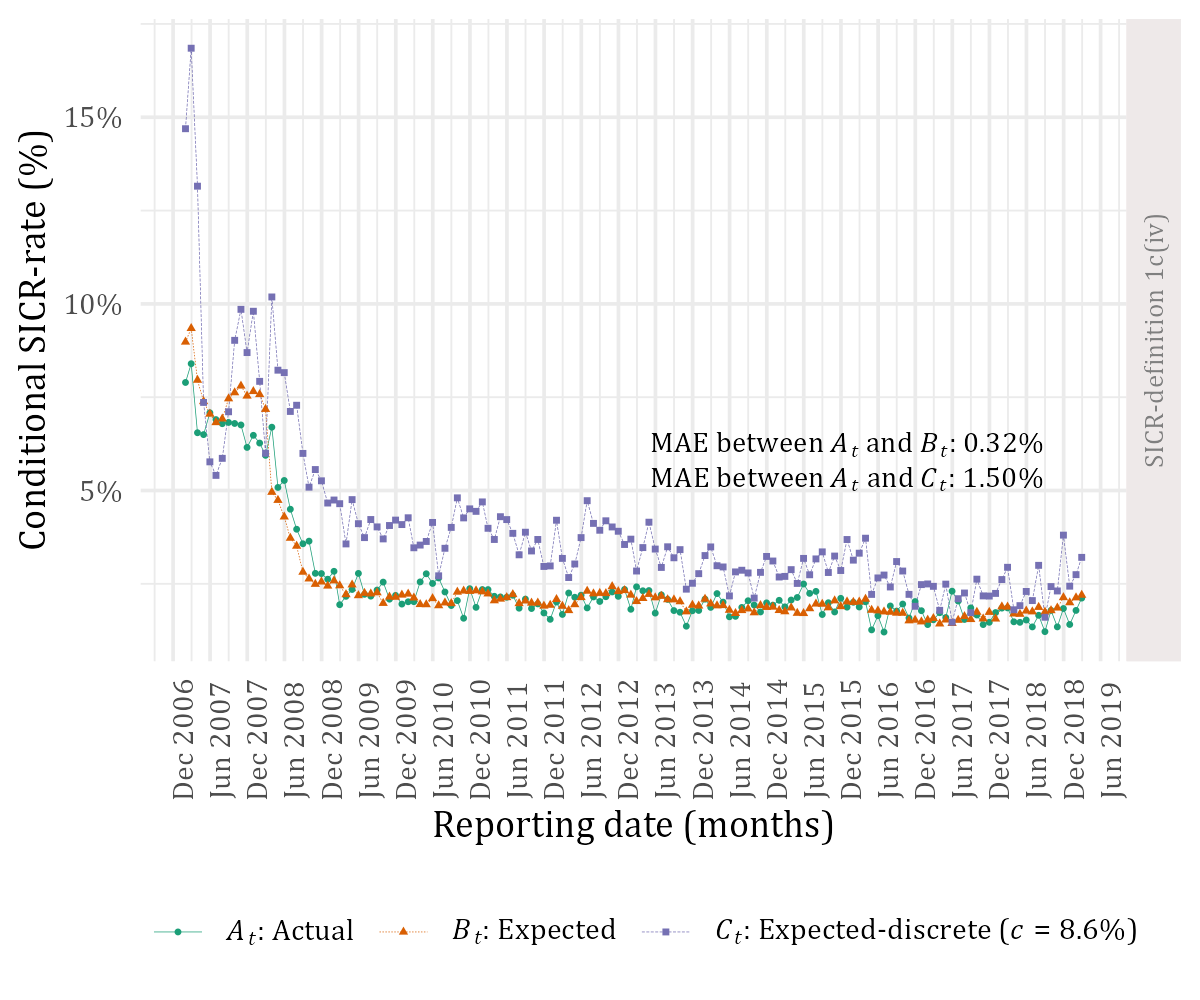}\label{fig:1c-SICR-Inc_d}
\end{subfigure}
\caption{Time graphs of actual versus expected SICR-rates within $\mathcal{D}_S$ for SICR-definition class 1c ($d=1,s=3$) across outcome periods $k\in [3,6,9,12]$. Graph design follows that of \autoref{fig:1a_SICR_Rates}.}\label{fig:1c-SICR-Inc}
\end{figure}

As in \autoref{fig:1a_SICR_Rates}, we compare the time graphs of $A_t$ versus $B_t$ across both $k\in\{3,6,9,12\}$ and $s\in\{1,2,3\}$; see \crefrange{fig:1b_SICR_Rates}{fig:1c-SICR-Inc}. Having calculated the MAE $m_1$ between $A_t$ and $B_t$ for a given $(k,s)$-tuple, the average discrepancy is mostly similar across $k$-values when keeping $s$ constant. By implication, the aggregated predictions from the associated SICR-models agree closely with observed reality for any $k$, despite the growing stagnancy as $s$ increases in both $A_t$ and $B_t$.
Moreover, when calculating the mean MAE-value across $k$ for each $s$-value, i.e., $\{0.44\%,0.37\%,0.32\%\}$, the decreasing trend in error implies greater agreement with observed reality, as corroborated by the greater AUC-values within \crefrange{tab:1a-performance}{tab:1bc-performance}.
We repeat the same exercise in comparing $A_t$ with $C_t$, whereafter we obtain mean MAE-values of $\{1.03\%, 1.07\%, 1.22\%\}$ respective to each $s$-value. Being about 2-4 times greater than the previous set of means, it implies less agreement with observed reality, which is somewhat disappointing. However, the discretised expected SICR-rates are known to be highly sensitive to the choice of the cut-off $c_{dsk}$, which regrettably impedes any further analyses on $C_t$ vs $A_t$.
That said, and without changing the misclassification cost ratio $a$, larger $s$-values can still achieve a greater prevalence of $C_t\geq A_t$, i.e., `over-prediction', which is reassuringly risk-prudent under IFRS 9.

In summary, these results suggest the following prudential optima in defining SICR-events for $d=1$, given the trade-offs amongst AUC-values, dynamicity $\omega_{dsk}$, instability $\sigma_{dsk}$, and the responsiveness/resiliency of resulting SICR-rates amidst macroeconomic malaise. For no stickiness $s=1$, choose $k\in[6,12]$; for some stickiness $s=2$, choose $k\in[6,9]$; and for a large degree of stickiness $s=3$, choose $k=9$. Smaller $k$-values within these ranges will yield more accurate and dynamic SICR-predictions at the account-level. These benefits are traded for portfolio-level SICR-rates that will become less dynamic, have lower means, and are less responsive to externalities. Larger $s$-values will also produce more accurate but less dynamic SICR-predictions, though the resulting SICR-rates have markedly lower means and are increasingly insensitive to externalities due to their growing stagnancy/stability over time. These trade-offs are intuitively balanced when choosing $k=9$ across $s$ as well as at $s=2$ across $k\in\{6,9\}$.

\subsection{The negative impact of greater delinquency \texorpdfstring{$d=2$}{Lg} within SICR-definitions}
\label{sec:delinquency}

From \S5.5.11 and \S B5.5.20 of IFRS 9, a SICR-outcome is said to occur once the arrears has reached \textit{30 days past due}, i.e., $g_0(t)\geq d$ where $d=1$. However, this presumption (or `backstop') can be rebutted if there is evidence against the supposed deterioration of credit quality, despite the delinquency accruing to $g_0(t)=2$. 
We therefore fixed $d=2$ within our SICR-framework, thereby resulting in the remaining twelve SICR-definitions and associated SICR-models, i.e., classes 2a-c in \autoref{tab:SICR_Defs} (darker shades). 
Reassuringly, \autoref{tab:AllDef_performance} shows that the high-level trends remain largely intact (albeit greatly muted) for most of the performance measures across $k$ and $s$ for $d=2$, at least compared to the results from \crefrange{sec:outcomePeriods}{sec:stickiness} for $d=1$. 
Under IFRS 9, the extremely low $\phi_{2sk}$-rates imply that the resulting Stage 2 provisions would be similarly small, which is unintuitive given the greater underlying delinquency associated with $d=2$. 
It is also questionable to build bespoke SICR-models with the `narrower' $d=2$ class when it is already contained within the `broader' $d=1$ class by definition.
Furthermore, the actual SICR-rates $A_t$ resulting from $d=2$ are significantly lower than those from $d=1$, i.e., $A_t(d=2,s,k) < A_t(d=1,s,k)$ over time $t$ and across all $(s,k)$-combinations. These rates are even lower than the default rates prevailing during the 2008-GFC, which contradicts IFRS 9 in providing \textit{timeously} for credit losses. Moreover, and given $\sigma_{dsk}$, these $A_t(2,s,k)$-rates are not as dynamic as their $A_t(1,s,k)$-counterparts in responding to macroeconomic crises. 
Given these results, we recommend against using $d=2$ and therefore support using the backstop, as implicitly included within our SICR-framework when setting $d=1$.

\begin{table}[ht!]
\caption{Selected performance measures for evaluating SICR-models across all SICR-definitions from \autoref{tab:SICR_Defs}. Table design follows that of \autoref{tab:1a-performance}.}
\centering
\begin{subtable}{.47\linewidth}
    \label{tab:AllDef_performance_1}
    \centering
    \caption{Delinquency threshold $d=1$}
    \begin{tabular}{p{1.35cm} p{1.5cm}  p{1.4cm} p{2.1cm}}
    \toprule
    \textbf{Definition} &  \textbf{Prevalence} $\phi_{dsk}$ & \textbf{Instability} $\sigma_{dsk}$ & \textbf{AUC-Probabilistic} \\ \midrule
    \rowcolor[HTML]{ECF4FF} 
    1a(i) & 6.16\% & 1.00\% & 91.3\% \footnotesize{$\pm$ 0.48\%} \\
    \rowcolor[HTML]{ECF4FF} 
    1a(ii) & 6.13\% & 1.43\% & 88.5\% \footnotesize{$\pm$ 0.54\%} \\
    \rowcolor[HTML]{ECF4FF} 
    1a(iii) & 6.07\% & 1.64\% & 86.5\% \footnotesize{$\pm$ 0.57\%} \\
    \rowcolor[HTML]{ECF4FF} 
    1a(iv) & 5.99\% & 1.81\% & 84.8\% \footnotesize{$\pm$ 0.60\%}\\
    \rowcolor[HTML]{E6FFE6} 
    1b(i) & 4.74\% & 1.07\% & 93.8\% \footnotesize{$\pm$ 0.43\%} \\
    \rowcolor[HTML]{E6FFE6} 
    1b(ii) & 4.72\% & 1.38\% & 89.4\% \footnotesize{$\pm$ 0.56\%} \\
    \rowcolor[HTML]{E6FFE6} 
    1b(iii) & 4.68\% & 1.54\% & 88.0\% \footnotesize{$\pm$ 0.59\%} \\
    \rowcolor[HTML]{E6FFE6} 
    1b(iv) & 4.61\% & 1.71\% & 86.5\% \footnotesize{$\pm$ 0.63\%} \\
    \rowcolor[HTML]{FFF2E2} 
    1c(i) & 3.82\% & 0.97\% & 95.7\% \footnotesize{$\pm$ 0.38\%} \\
    \rowcolor[HTML]{FFF2E2} 
    1c(ii) & 3.81\% & 1.28\% & 91.6\% \footnotesize{$\pm$ 0.57\%}\\
    \rowcolor[HTML]{FFF2E2} 
    1c(iii) & 3.78\% & 1.44\% & 88.9\% \footnotesize{$\pm$ 0.64\%} \\ 
    \rowcolor[HTML]{FFF2E2} 
    1c(iv) & 3.73\% & 1.57\% & 86.7\% \footnotesize{$\pm$ 0.70\%} \\ \bottomrule
    \end{tabular}
\end{subtable}
\begin{subtable}{.47\linewidth}
    \label{tab:AllDef_performance_2}
    \centering
    \caption{Delinquency threshold $d=2$}
    \begin{tabular}{p{1.35cm} p{1.5cm}  p{1.4cm} p{2.1cm}}
    \toprule
    \textbf{Definition} &  \textbf{Prevalence} $\phi_{dsk}$ & \textbf{Instability} $\sigma_{dsk}$ & \textbf{AUC-Probabilistic} \\ \midrule
    \rowcolor[HTML]{C0DAFE} 
    2a(i) & 0.64\% & 0.23\% & 91.8\% \footnotesize{$\pm$ 1.15\%} \\
    \rowcolor[HTML]{C0DAFE} 
    2a(ii) & 0.64\% & 0.26\% & 86./\% \footnotesize{$\pm$ 1.57\%} \\
    \rowcolor[HTML]{C0DAFE} 
    2a(iii) & 0.64\% & 0.26\% & 84.6\% \footnotesize{$\pm$ 1.64\%} \\
    \rowcolor[HTML]{C0DAFE} 
    2a(iv) & 0.63\% & 0.26\% & 81.7\% \footnotesize{$\pm$ 1.86\%}\\
    \rowcolor[HTML]{B5FFB5} 
    2b(i) & 0.20\% & 0.10\% & 97.2\% \footnotesize{$\pm$ 1.08\%} \\
    \rowcolor[HTML]{B5FFB5} 
    2b(ii) & 0.20\% & 0.11\% & 90.5\% \footnotesize{$\pm$ 2.35\%} \\
    \rowcolor[HTML]{B5FFB5} 
    2b(iii) & 0.20\% & 0.11\% & 89.9\% \footnotesize{$\pm$ 2.28\%} \\
    \rowcolor[HTML]{B5FFB5} 
    2b(iv) & 0.20\% & 0.11\% & 83.5\% \footnotesize{$\pm$ 3.01\%} \\
    \rowcolor[HTML]{FFE1BD} 
    2c(i) & 0.07\% & 0.06\% & 89.8\% \footnotesize{$\pm$ 3.57\%} \\
    \rowcolor[HTML]{FFE1BD} 
    2c(ii) & 0.07\% & 0.06\% & 78.4\% \footnotesize{$\pm$ 4.56\%}\\
    \rowcolor[HTML]{FFE1BD} 
    2c(iii) & 0.07\% & 0.06\% & 71.3\% \footnotesize{$\pm$ 4.83\%} \\ 
    \rowcolor[HTML]{FFE1BD} 
    2c(iv) & 0.08\% & 0.06\% & 70.1\% \footnotesize{$\pm$ 4.61\%} \\ \bottomrule
    \end{tabular}
\end{subtable}
\label{tab:AllDef_performance}
\end{table}


\subsection{Comparing approaches: SICR-modelling vs PD-comparison}
\label{sec:approach_comparison}

Our SICR-modelling framework aims to provide a new and flexible way of conducting SICR-classification that is more proactive, focused, accurate, and dynamic than that of the classical PD-comparison approach. It is only natural then to compare the old to the new in identifying a superior approach.
As initially described in \autoref{sec:background}, the PD-comparison approach relies on calculating the change (or \textit{magnitude}) between two risk estimates of an account at two different time points. This magnitude $m$ is then evaluated against a chosen threshold $u>0$ in classifying a loan into either Stage 1 or 2 of credit impairment. Given any $u$-value, we subsequently formulate this approach into the binary-valued decision model $\mathcal{H}(m,u) \in \{0,1\}$, as applied on each account over its lifetime, and defined using Iverson brackets $[\cdot]$ as
\begin{equation} \label{eq:pd_comparison_decision}
    \mathcal{H}(m,u) = \left[ m > u\right] \ .
\end{equation}
This $m$-quantity can be substituted with the PD-ratio, which relates the PD-estimate of account $i$ at two different points over its lifetime, having used the basic PD-model from \autoref{app:PD_Model} to produce these PD-estimates. More formally, let $m(\boldsymbol{x}_{it},t)$ denote this PD-ratio for account $i$ given multivariate input data $\boldsymbol{x}_{it}$, as measured at each point $t=t_1,\dots,T_i-1$ from the account's initial recognition $t_1$ up to the penultimate period of its observed lifetime $T_i$. Having substituted $m(\boldsymbol{x}_{it},t)$ into \autoref{eq:pd_comparison_decision}, we obtain the discrete `prediction' $h(\boldsymbol{x}_{it})$ of the future SICR-status at $t+1$. In fact, we reuse our SICR-framework from \autoref{sec:method} by applying the $\mathcal{Z}_t(d=1,s=1,k=1)$-process from \autoref{eq:decision_rule_generator}, thereby creating the SICR-outcomes $y_{it}$ across all accounts $i$ and over their lifetimes $t$.
Accordingly, the actual and discretised expected 1-month SICR-rates can be estimated respectively from the resulting sample of $y_{it}$ and $h(\boldsymbol{x}_{it})$ values, followed by their comparison.

\begin{figure}[ht!]
\centering\includegraphics[width=0.7\linewidth,height=0.52\textheight]{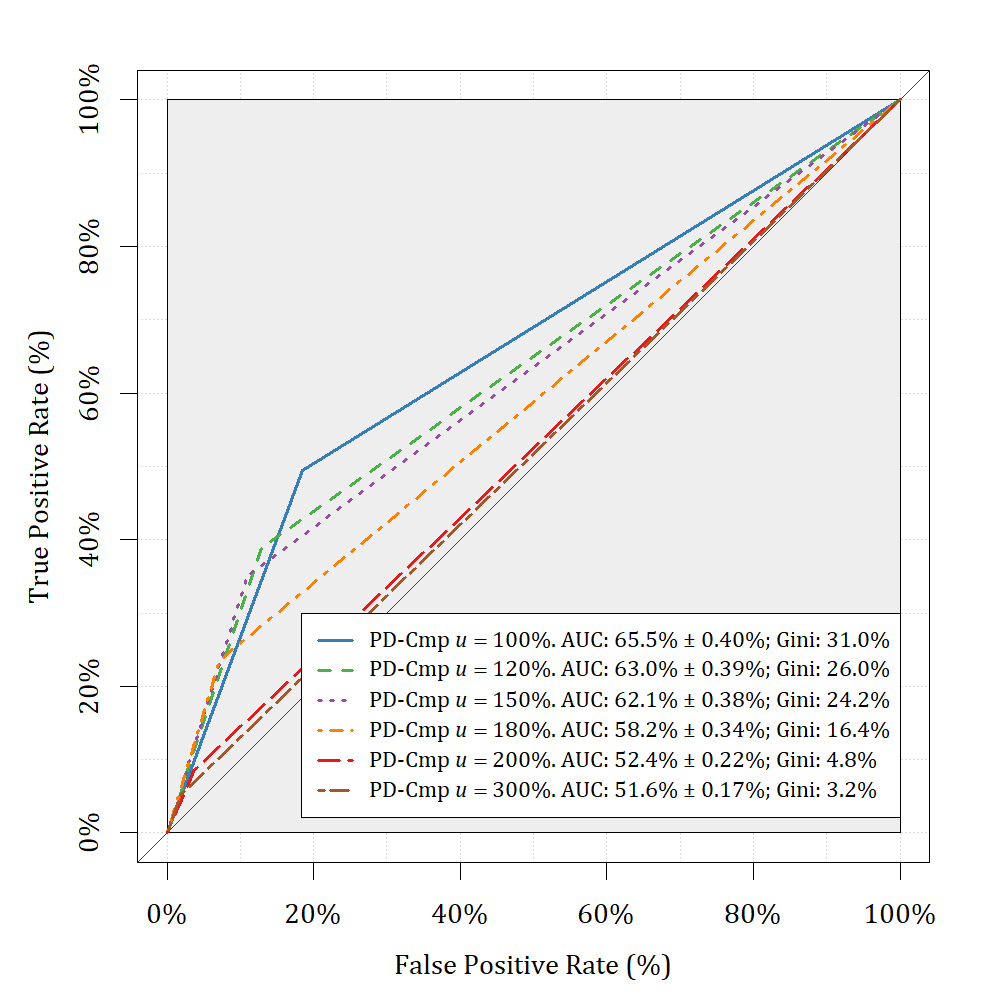}
\caption{ROC-analysis of the decision model $\mathcal{H}$ from \autoref{eq:pd_comparison_decision}, which is used as a discrete classifier across candidate $u$-thresholds in following the PD-comparison approach. The AUC-values are printed using $\mathcal{D}_V$, together with 95\% confidence intervals (DeLong-method) and corresponding Gini-values.}\label{fig:PD-Comp-ROC}
\end{figure}

As discussed in \autoref{sec:background}, choosing an appropriate $u$-value in order to use the decision model $\mathcal{H}$ from \autoref{eq:pd_comparison_decision} is a subjective and contentious task in practice. We therefore examine a few different choices of $u\in \{100\%, 120\%, 150\%, 180\%, 200\%, 300\% \}$ using discretion, though which deliberately includes the candidate $u=200\%$ from the \citet{EBA_stress2018}, or EBA.
The prediction accuracy of the resulting SICR-classification is similarly gauged using ROC-analysis in the validation set $\mathcal{D}_V$, as shown in \autoref{fig:PD-Comp-ROC} for each $u$-value. In summarising the ROC-analysis, the AUC-values clearly indicate that the prediction accuracy of $\mathcal{H}$ is substantially inferior to that of any SICR-model in \crefrange{tab:1a-performance}{tab:1bc-performance}, regardless of $u$. Moreover, the AUC-values appear to be a monotonically decreasing function of $u$, where the AUC seems to deteriorate rapidly for $u\geq150\%$.
The EBA-recommended threshold of $u=200\%$ produced some of the most inaccurate SICR-predictions. However, this result is at least partially dependent on the quality of the underlying PD-model; itself deemed typical within the present context, as argued in \autoref{app:PD_Model}.
In parametrising $\mathcal{H}$, we therefore select the $u$-threshold that yielded the greatest AUC-value, as well as the EBA-threshold; i.e., $u\in\{100\%, 200\%\}$. Both variants of $\mathcal{H}$ can fairly represent the PD-comparison approach from at least two perspectives: 1) maximising prediction accuracy; and 2) adhering to regulatory prescription, albeit sub-optimal.

\begin{figure}[ht!]
\centering\includegraphics[width=0.78\linewidth,height=0.71\textheight]{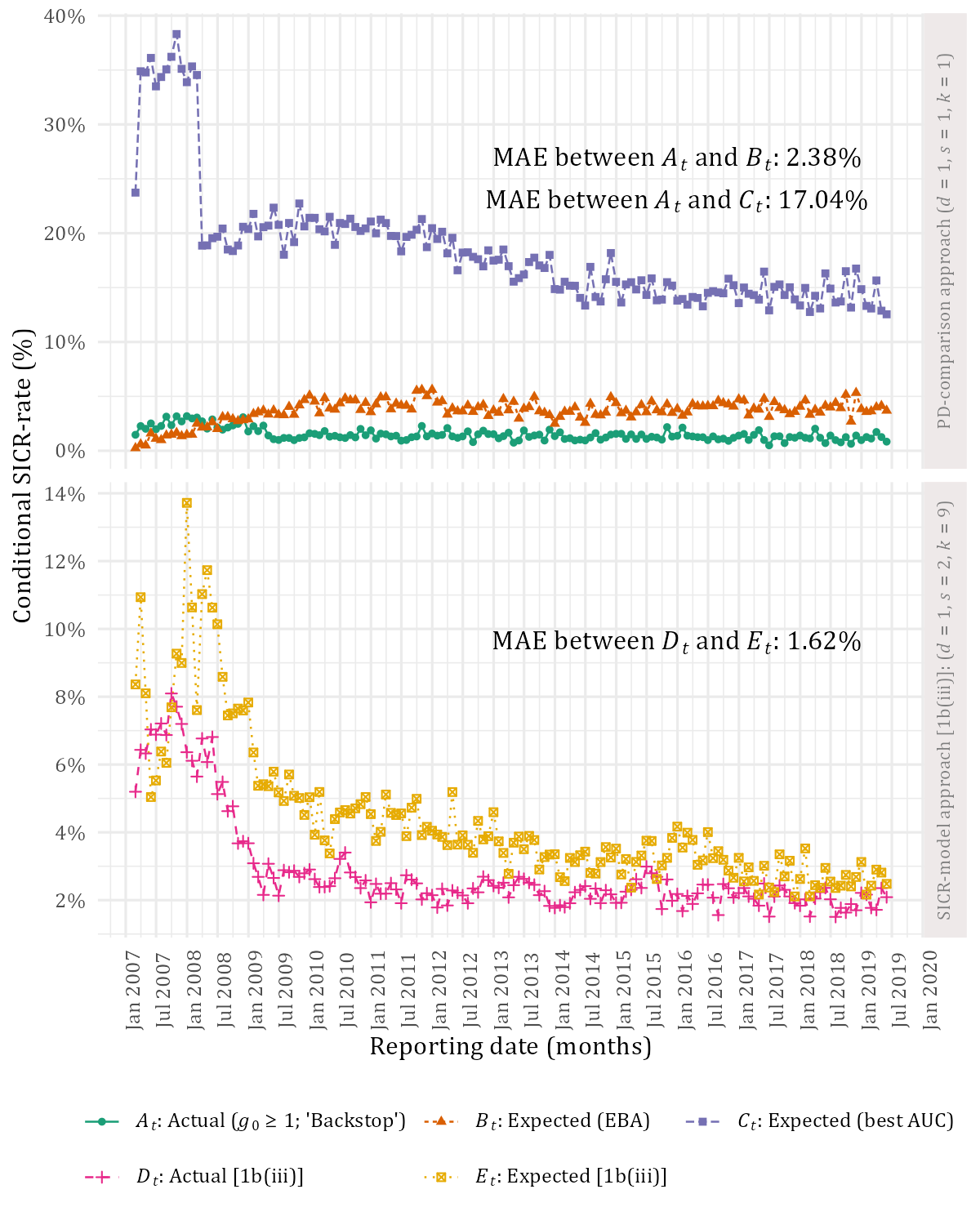}
\caption{Time graphs of actual vs discretised expected SICR-rates, respective to following the PD-comparison approach vs the SICR-modelling approach with definition 1b(iii). The PD-comparison approach includes two candidate $u$-thresholds for parametrising the decision model $\mathcal{H}$ in \autoref{eq:pd_comparison_decision}: 1) $u=200\%$ from the \citet{EBA_stress2018}; 2) $u=100\%$ that yielded the best AUC-value. Graph design follows that of \autoref{fig:1a_SICR_Rates}. }\label{fig:SICR_Rates-ApproachComp}
\end{figure}

In comparing approaches, we select SICR-definition 1b(iii) from \autoref{tab:SICR_Defs} and consider the resulting SICR-model, which was previously motivated in \autoref{sec:stickiness} as one of the best-performing SICR-models. 
Using the corresponding $c_{129}$-value from \autoref{tab:1bc-performance} for this definition $(d=1,s=2,k=9)$, we dichotomise this SICR-model and similarly evaluate its discrete predictions within the validation set $\mathcal{D}_V$ using ROC-analysis. The resulting AUC-value of 76.8\% indicates a decent level of accuracy, which compares favourably to that of the $\mathcal{H}$-classifier. In particular, the PD-comparison approach yielded lower AUC-values of 66.36\% and 52.44\% respective to $u=\{100\%, 200\%\}$.
While its account-level prediction accuracy is clearly atrocious, the $\mathcal{H}$-classifier may perform more admirably on the portfolio-level. We therefore compare the time graphs of actual vs discretised expected SICR-rates, respective to both approaches and shown in \autoref{fig:SICR_Rates-ApproachComp}.
All of the expected rates across both approaches seemingly exceed their actual counterpart for most periods, which is certainly risk-prudent under IFRS 9. However, the degree of such over-prediction amounts to misallocated funds and wasted provisions, which can again be measured using the MAE between two SICR-rates. The EBA-threshold ($u=200\%$) achieves a respectable MAE-value, which is reasonably close to that of the SICR-model, albeit still worse. More importantly, the MAE-value of the EBA-threshold is more than 7 times lower than that of the best-AUC threshold ($u=100\%$). Despite its improved prediction accuracy at the account-level, the best-AUC threshold clearly results in an overly conservative SICR-rate at the portfolio-level, which surely poses an immense opportunity cost.

\autoref{fig:SICR_Rates-ApproachComp} also demonstrates different volatility patterns in the underlying SICR-rates across both approaches. 
In particular, the actual SICR-rate (in green) from the PD-comparison approach reacted rather mildly to the onset of the 2008-GFC, relative to its counterpart (in pink) from the SICR-modelling approach. We largely ascribe this result to the former's use of $k=1$, which usually causes volatility in PD-modelling due to risk immaturity in the outcomes; see \citet{kennedy2013window} and \citet{mushava2018experimental}. 
On the other hand, the two expected SICR-rates from the PD-comparison approach either under-predict their actual counterpart during the 2008-GFC (shown in orange: EBA-threshold), or massively over-predict it (shown in purple: best-AUC threshold). Both of these results are unsatisfactory and highlight the main drawbacks of the PD-comparison approach: 1) its tacit reliance on an appropriately accurate PD-model; and 2) its extraordinary sensitivity to the $u$-threshold.
In contrast, both rates (in yellow and pink) from the SICR-modelling approach react more flexibly and intuitively as the 2008-GFC unfolds, having achieved a more pronounced peak at the height of the crisis without becoming excessive. Although decent, the dichotomisation of the SICR-model can surely be improved in future studies by tweaking the $c_{129}$-threshold, which should result in even better performance. 
However, and in finalising the approach comparison, the evidence suggests that the SICR-modelling approach is objectively the superior approach.

\section{Conclusion}
\label{sec:conclusion}

The meaning of a SICR-event has become needlessly nebulous when modelling loan impairments under IFRS 9. The resulting complexity is arguably a consequence of using an approach based on drawing arbitrary PD-comparisons; an approach with at least two prominent challenges. 
Firstly, it requires PD-estimates that are reasonably accurate at any two time points, which is itself challenging. Secondly, the approach requires evaluating the difference between any two PD-estimates against a subjectively-chosen threshold, whose selection can be ambiguous and contentious. 
Intuitively, too small a threshold can trigger the mass migration of loans into Stage 2, which can become prohibitively cost-inefficient and overly conservative. On the other hand, too large a threshold may never be materially breached, thereby keeping loans naively in Stage 1 and leaving a bank grossly under-provided.
At the moment, choosing any threshold is non-trivial given the lack of an overarching optimisation framework. Practitioners and regulators alike have little choice but to rely on subjective discretion and/or regulatory prescription; both of which can be sub-optimal. More generally, these two challenges of the PD-comparison approach can counteract the main imperative of IFRS 9, i.e., recognising credit losses timeously.

As an alternative, we contribute a concise and simple SICR-framework from which SICR-definitions may be generated and tested. Any such (target) definition can then be used in building a statistical SICR-model (or supervised binary classifier), which is premised on predicting the probability of future delinquency for non-delinquent accounts; itself another contribution. This SICR-model can probabilistically classify a performing loan into either Stage 1 or 2, using a rich and dynamic set of macroeconomic and obligor-specific input variables. As supported by \S B.5.5.12 in IFRS 9, our SICR-modelling approach does not rely on PD-comparisons and therefore requires neither underlying PD-models nor selecting any related threshold. 
Our approach is more parsimonious than PD-comparisons since the inputs of a SICR-model can relate more directly to the \textit{change} in delinquency risk, instead of just default risk alone. As one of these inputs, the PD-ratio already signifies the change in risk since initial recognition, thereby rendering the predictions from a SICR-model as compliant with \S5.5.9 of IFRS 9.
Moreover, a SICR-modelling approach allows drawing statistical inference on the drivers of the overall SICR-process as a stochastic phenomenon, which can certainly help in portfolio management.
Lastly, our approach prevents any pre-existing issues within a PD-model from bleeding into staged impairment classification under IFRS 9, which can be another practical benefit.

In generating SICR-definitions, our framework avails three useful parameters: 1) the delinquency threshold $d$ in testing accrued delinquency at any point; 2) the level of stickiness $s$ when testing delinquency over consecutive periods; and 3) the outcome period $k$ over which to predict SICR-statuses. 
In varying these parameters, we effectively produced 27 different SICR-definitions as unique combinations of the triple $(d,s,k)$. Each SICR-definition is applied on the same South African mortgage data from 2007-2019, whereupon an account-level SICR-model is estimated using binary logistic regression per definition. 
We demonstrate that shorter outcome periods can yield SICR-predictions that are increasingly more accurate and dynamic over loan life, at least for $k\geq 6$ months. However, upon aggregating these account-level predictions to the portfolio-level, the resulting SICR-rate appears less dynamic over time for smaller $k$-values, have progressively lower means, and are increasingly insensitive to unfolding economic crises like the 2008-GFC. Some of these relationships are not necessarily linear: overly long outcome periods ($k\geq 18$) yield SICR-rates that are similarly unresponsive to market failures, in addition to the degrading prediction accuracy.

The $s$-parameter has a stabilising yet costly effect on SICR-classification, wherein SICR-events become scarcer as $s$ increases. Greater stickiness yield account-level SICR-predictions that are more accurate but also less dynamic over loan life. From these stickier SICR-definitions, the resulting portfolio-level SICR-rates become less dynamic over time, have lower means, and are increasingly insensitive to the 2008-GFC.
Furthermore, both $k$ and $s$ parameters interact with each other in that SICR-predictions become more accurate as $k$ decreases and $s$ increases. However, the dynamicity of account-level predictions decreases for larger $s$ but increases again for smaller $k$. 
Lastly, choosing $d=2$ yields extremely scarce SICR-events across all values of $s$ and $k$, which would compromise the resulting Stage 2 provision-levels if used; a result that supports the `backstop' ($d=1$) of IFRS 9.
These trends form a reusable analytical framework in which any SICR-definition can be examined on the following four factors: the accuracy and dynamicity of the resulting SICR-predictions, the instability of implied SICR-rates, and its responsiveness to economic distress. 
A reasonable trade-off exists amongst these factors when choosing $k=9$ across any $s$-value, as well as when selecting $s=2$ across $k\in\{6,9\}$.
Having selected the best-performing SICR-model with definition $(d=1,s=2,k=9)$, we deliberately compare its predictions to those yielded by the PD-comparison approach. Not only does the latter approach result in a worse prediction accuracy, but it also leads to expected SICR-rates that are either too conservative during normal times, or insufficiently so during economic crises.
Our SICR-modelling approach can therefore yield predictions that are both highly accurate and reasonably dynamic over time, while the resulting SICR-rates remain relatively stable though still reassuringly sensitive to externalities.

Future research can examine SICR-modelling using data from other loan portfolios and across other credit markets; both of which may affect the resulting choices of $(k,s)$. In this regard, future studies can explore an even finer-grained list of $k$-values, particularly for $k\in[4,12]$. Doing so can refine the relationships that we have found, which can help in devising an optimisation framework for selecting parameters optimally. Alternatively, the $k$-parameter can be embedded more dynamically within a broader survival modelling approach, thereby enabling SICR-prediction across any number of $k$-values over loan life. Doing so can enable SICR-prediction across any number of $k$-values over loan life, instead of just the 7 fixed $k$-values that we have explored.
As for modelling techniques in general, future researchers can certainly expand our study by experimenting with more advanced binary classifiers than logistic regression, e.g., Support Vector Machines as in \citet{harris2013b}. 
As another avenue, a future study might focus on stress-testing any relevant input variables (e.g., macroeconomic covariates) within a SICR-model, perhaps towards forecasting overall SICR-rates given a particular macroeconomic scenario.

The misclassification cost ratio $a$ within the Generalised Youden Index $J_a$ can (and should) be tweaked towards improving the discrete output of a probabilistic SICR-model. Doing so would amount to changing (indirectly) the cut-off probability beyond which a SICR-event is predicted. If this cut-off is inappropriately chosen without any analysis, then the same criticism applies that we have made about a subjectively-chosen threshold within the PD-comparison approach. I.e., an inappropriate cut-off would introduce bias into overall SICR-classification; though the degree thereof is unstudied, which certainly warrants further research.
Relatedly, one might experiment with using $J_a$ towards finding a suitable cut-off within the PD-comparison approach. Lastly, future researchers can embed the misclassification cost itself when training a SICR-model, perhaps using a bespoke loss function, instead of imposing such costs exogenously and exterior to the model afterwards.


\subsection*{Acknowledgements}
\noindent This work is financially supported in part by a National Research Foundation grant, with no known conflicts of interest that may have influenced the outcome of this work. The authors would like to thank Prof. Dirk Tasche for valuable remarks on our work, as well as all anonymous referees and editors for their extremely valuable contributions that have substantially improved this work.

\appendix
\section{Appendix}
\label{app:app01}

The design of the resampling scheme is described in \cref{app:samplingDesign}, followed by testing and verifying its representativeness across all resulting samples.
In \cref{app:logistic}, we discuss the fundamentals of a statistical technique called \textit{binary logistic regression}, its use in quantitative finance, as well as the Generalised Youden Index $J_a$ in dichotomising a logit-model. 
Thereafter, the interactive process by which input variables are selected is briefly discussed in \cref{app:inputSpace}, followed by summarising the input spaces across all SICR-models. We also discuss some interesting patterns that were found during feature selection.
A basic PD-model is described in \cref{app:PD_Model} for generating an important input variable (\verb|PD-ratio|) that was used across all SICR-models, thereby achieving full compliance with IFRS 9.
Finally, we define a few performance measures in \cref{app:PerfMeasures} for assessing various aspects of our modelling results.

\subsection{The design and testing of the resampling scheme regarding its representativeness}
\label{app:samplingDesign}

\begin{figure}[ht!]
\centering\includegraphics[width=0.9\linewidth,height=0.55\textheight]{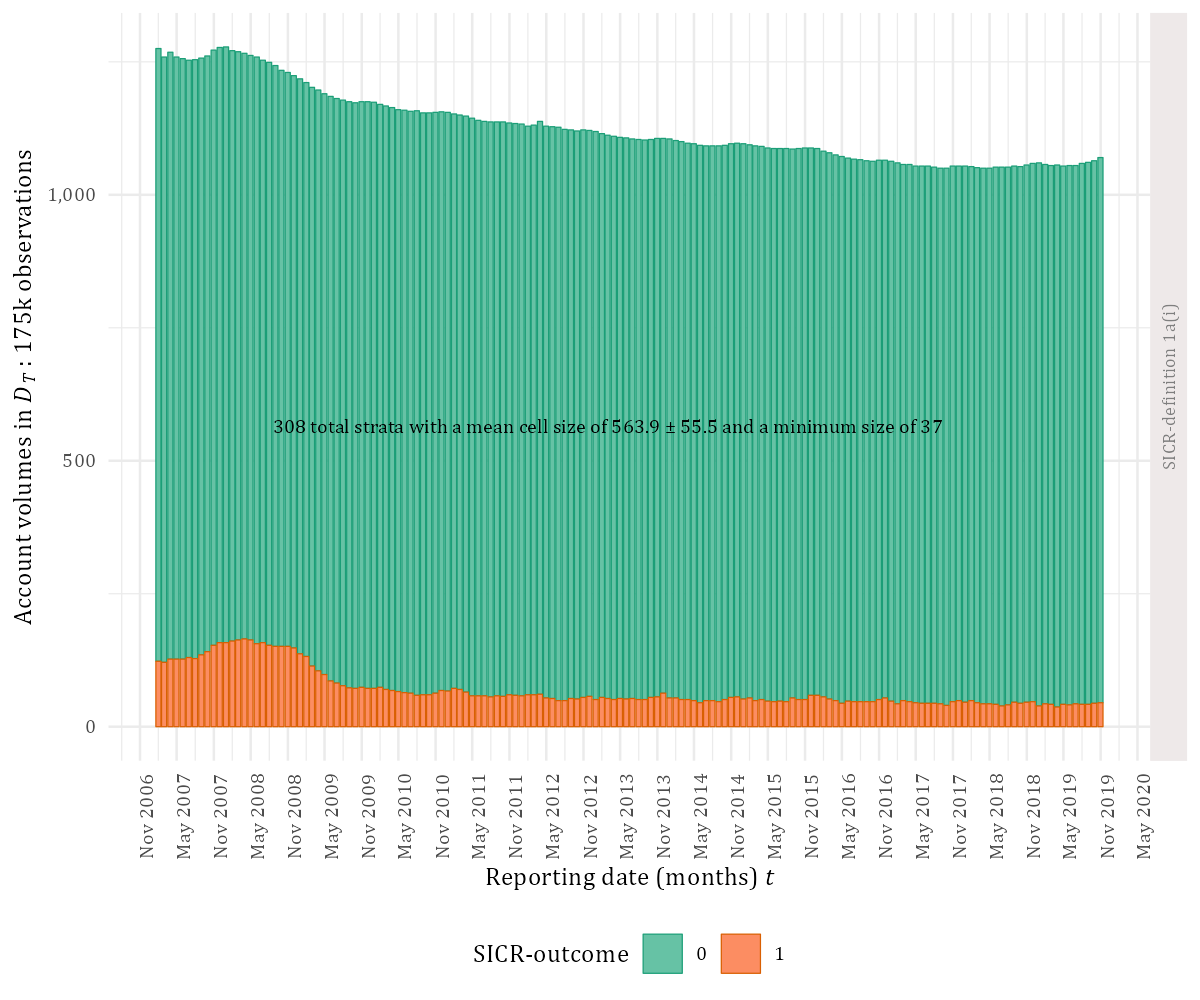}
\caption{Subsampled account volumes are shown over time and grouped by SICR-outcome, having used the 1a(i)-definition from \autoref{tab:SICR_Defs}; itself applied on $\mathcal{D}_T$. Both time and the SICR-outcome are stratifiers within a two-way stratified resampling scheme. Summary statistics are overlaid with a 95\% confidence interval.}\label{fig:SICR_Strata_analysis}
\end{figure}

\begin{figure}[ht!]
\centering\includegraphics[width=0.9\linewidth,height=0.55\textheight]{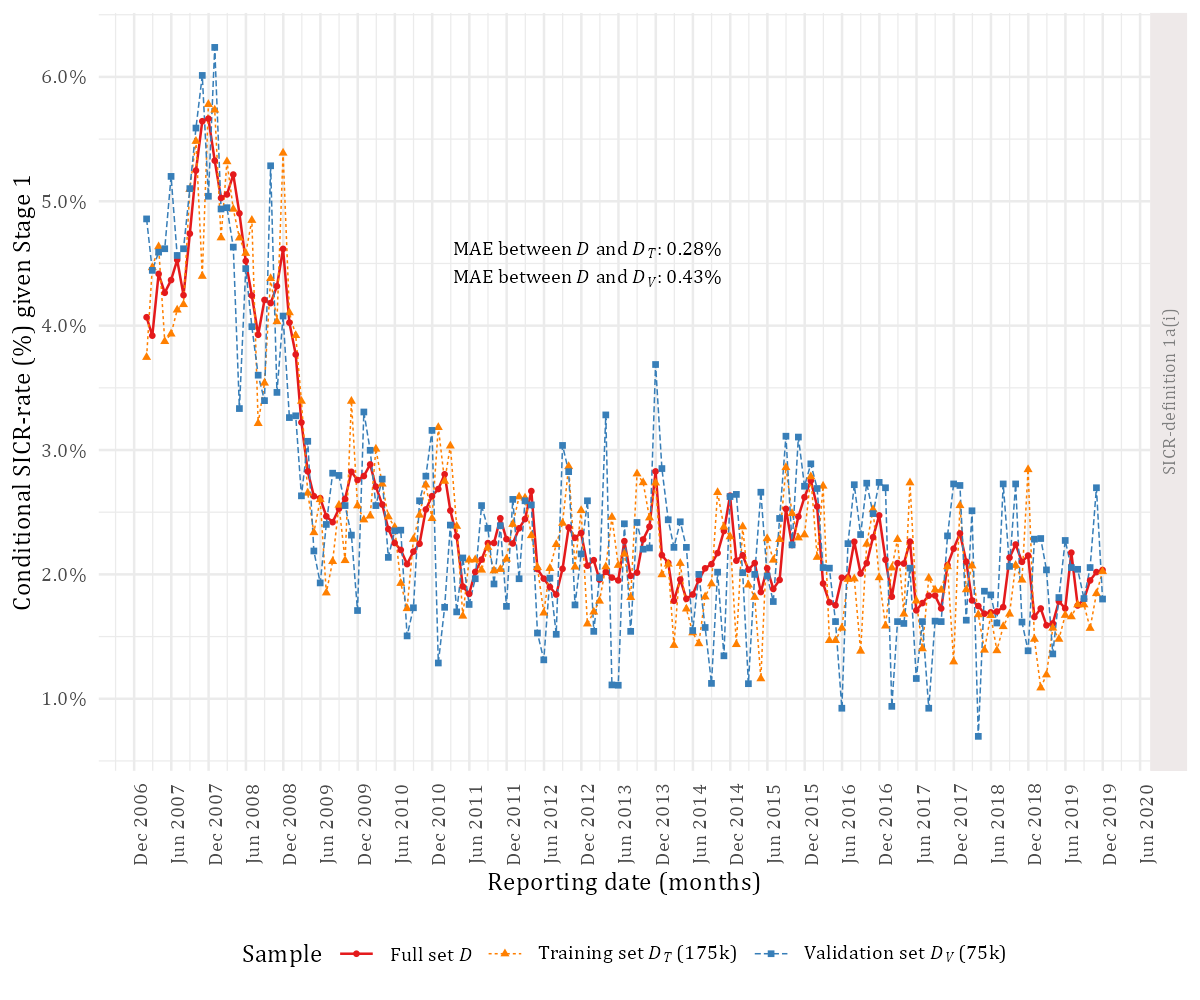}
\caption{Comparing actual SICR-rates over time across the various datasets, having used the 1a(i)-definition from \autoref{tab:SICR_Defs}. The Mean Absolute Error (MAE) between each sample and the full set $\mathcal{D}$ is overlaid in summarising the line graph discrepancies over time. }\label{fig:SICR_Incidence_Samples}
\end{figure}

For every SICR-definition in \autoref{tab:SICR_Defs}, the raw dataset $\mathcal{D}$ is grouped by the binary-valued SICR-outcomes that result from applying the $\mathcal{Z}_t(d,s,k)$-process from \autoref{eq:decision_rule_generator} within each monthly cohort $t$, thereby resulting in about 310 strata. Observations are then sampled randomly within each stratum in creating the sub-sampled dataset $\mathcal{D}_S$. The sampling proportion is dynamically set for each SICR-definition such that each $\mathcal{D}_S$ will be of the same fixed size, i.e., about 250,000 monthly observations in total. Finally, a simple cross-validation resampling scheme is used (with a 70\%-30\% ratio) to partition the data $\mathcal{D}_S$ into two non-overlapping sets: a training set $\mathcal{D}_T$ and a validation set $\mathcal{D}_V$; see \citet[pp.~249--254]{hastie2009elements}. 
Considering the 1a(i)-definition from \autoref{tab:SICR_Defs}, we graph the account volumes over time in \autoref{fig:SICR_Strata_analysis} for one such $\mathcal{D}_T$-set, thereby affirming its adequacy for statistical modelling.
In verifying the sampling representativeness, we compare the actual SICR-rates (as defined in \autoref{eq:SICR_Rate}) over time and across the resampling scheme, as illustrated in \autoref{fig:SICR_Incidence_Samples}.
Evidently, the line graphs are reasonably close to one another across all samples. Furthermore, the Mean Absolute Error (MAE) of the SICR-rates between $\mathcal{D}$ and each respective sample is calculated as  $\mathcal{D}_T$: 0.28\% and $\mathcal{D}_V$: 0.43\%
; both of which are deemed as reasonably low. Similar results hold for all other SICR-definitions, which suggests that the resampling scheme is indeed representative of the population at large. 

\subsection{Dichotomising a binary logistic regression model using the Generalised Youden Index \texorpdfstring{$J_a$}{Lg}}
\label{app:logistic}

The literature on using binary logistic regression (LR) as a generic supervised classifier is quite extensive; see \citet[pp.~1--10]{hosmer2000logistic}, \citet[\S4]{bishop2006pattern}, \citet[\S4]{hastie2009elements}, and \citet[\S4.3]{james2013introduction}. 
Furthermore, its use within banking is ubiquitous, particularly in the field of application credit scoring, as was first demonstrated in \citet{wiginton1980}. The technique is considered by many authors to be the most successful modelling technique thus far in quantitative finance, as discussed by \citet{hand1997statistical}, \citet[\S4.5,~\S10--11]{thomas2002credit}, \citet{siddiqi2005credit}, \citet[\S1]{thomas2009consumer}, and \citet{bolton2010logistic}. Beyond application credit scoring, this technique is also typically used in pre-screening loan offers, detecting fraud cases, scoring collection success, informing direct marketing offers, and in risk-based pricing. As such, the ubiquity of the logistic regression technique suggests its use in the present study. At the very least, this technique and its results can serve as a benchmark when using more advanced classification techniques in future.

In dichotomising such an LR-model, consider its probability scores $p_1(\boldsymbol{x})$ given the inputs $\boldsymbol{x}$, which estimate the conditional probability of the positive event $\mathcal{C}_1$ given $\boldsymbol{x}$, i.e., $\mathbb{P}\left[\mathcal{C}_1 \vert \, \boldsymbol{x} \right]$.
These scores will need to be dichotomised in yielding binary 0/1-decisions, which implies choosing a cut-off $c\in [0,1]$ such that the discretised classifier $h(\boldsymbol{x})=1$ if $p_1(\boldsymbol{x}) > c$ and $h(\boldsymbol{x})=0$ if otherwise. For every possible $c$-value, the probability of a true positive (or SICR-event correctly predicted as such) is $q(c)=\mathbb{P}\left(p_1(\boldsymbol{x}) > c \, | \, \mathcal{C}_1\right)$, also known as sensitivity. Likewise, the probability of a true negative event $\mathcal{C}_0$ (or non-event correctly predicted as such) is $p(c)=\mathbb{P}\left(p_1(\boldsymbol{x}) \leq c \, | \, \mathcal{C}_0\right)$, also called specificity. 
In finding the optimal cut-off $c^*$ that incorporates both sensitivity and specificity, consider the Youden Index $J$ that is widely used in the biostatistical literature; see \citet{youden1950index}, \citet{greiner2000principles}, and \citet{schisterman2008youden}. This index $J$ is defined as the maximisation problem \begin{equation} \label{Youden_Index}
    J = \max_c{ \left\{q(c) + p(c) -1 \right\} } \, .
\end{equation} 
Clearly, the classical $J$ assigns equal weight to both sensitivity and specificity, which inappropriately equates the misclassification cost of a false negative to that of a false positive. However, and as shown by \citet{geisser1998comparing}, \citet{kaivanto2008maximization}, and \citet{schisterman2008youden}, the Generalised Youden Index $J_a$ improves upon $J$ by rendering $c$ sensitive to both types of misclassification costs. In particular, let $a>0$ be a cost multiple (or ratio) of a false negative relative to a false positive. If $\phi$ is the estimated prevalence of the $\mathcal{C}_1$-event, i.e., the prior $\mathbb{P}\left( \mathcal{C}_1 \right)$, then $J_a$ is expressed for a given $c$ as \begin{equation} \label{eq:gen_Youden_Index}
    J_a(c)= q(c) + \frac{1-\phi}{a\phi} \cdot p(c) - 1 \, ,
\end{equation} whereupon $c^*$ is given by \begin{equation} \label{eq:optimalCutoff}
    c^* = \arg \max_c{J_a(c)} \,.
\end{equation}

\subsection{Feature selection: constructing the input space of each SICR-model}
\label{app:inputSpace}

In finalising the input space of each SICR-model, we largely followed an interactive, multifaceted, and experimentally-driven process by which variables are selected across repeated logistic regressions.
One notable challenge to feature selection is that of large sample sizes, which are known to affect $p$-values when testing the statistical significance of regression coefficients, as demonstrated in \citet{lin2013LargeSamples}. The $p$-values can easily approach zero as the sample size increases, notwithstanding the greater statistical power availed by such larger sizes. This phenomenon overlaps with the \textit{Hughes principle} from \citet{hughes1968mean}: a model's predictive power will generally increase for every additional input, but decrease again after reaching some inflection point, provided that the sample size stays constant. 
Notwithstanding, and during our initial modelling attempts, we experimented with both a best subset approach (stepwise regression) and the LASSO shrinkage method in selecting inputs, as discussed in \citet[\S 6]{james2013introduction}. However, the necessary computation times proved to be excessive (especially for the stepwise method) and even unstable, whilst yielding negligible predictive performance and overly small models. 
Moreover, the declaration of \citet{henderson1981building} -- "the data analyst knows more than the computer" -- seems apt, cautioning against the practice of data dredging when automating feature selection, which is inevitably devoid of human expertise.

Our interactive selection process is guided by expert judgement, model parsimony, statistical significance, macroeconomic theory, and predictive performance on the validation set $\mathcal{D}_V$. The predictive performance is extensively evaluated using classical ROC-analysis, as summarised by the AUC-measure; see \citet{fawcett2006introduction}.  
In addressing the issue of large sample sizes, we model and select variables using a deliberately larger sub-sampled dataset $\mathcal{D}_S$ of 1 million observations, having applied the same resampling scheme from \cref{app:samplingDesign}. Once the selection process is concluded, the resulting input space is retested for statistical significance on a reduced subsampled set $\mathcal{D}_S$ that only contains 250,000 observations, from which the final SICR-model is also estimated.
We affirm that the vast majority of the inputs remain statistically significant across all $k$-values within each SICR-definition class in \autoref{tab:SICR_Defs}, which further reassures our selection (and `standardisation') process as robust. The interested reader can study our R-codebase on GitHub from \citet{botha2024sourcecode}, which details this process and the associated results as comments within the script of each SICR-model.

In \autoref{tab:featuresdescription}, we summarise and briefly describe the final set of input variables per SICR-definition class. 
Given their widespread prevalence, macroeconomic variables (and their lagged variants) have a notable impact on SICR-events irrespective of definition, which supports the forward-looking information requirement of IFRS 9 from \citet{ifrs9_2014}. 
Furthermore, the variable \verb|PD_ratio| signifies the change in the lifetime PD since initial recognition, which ensures compliance with \S5.5.9 of IFRS 9. Our results, however, show that this variable is statistically insignificant across all SICR-definitions, which implies that the broader input space already captures whatever intrinsic information this variable might have in predicting future SICR-events. 
This profound result clearly rebuts the underlying intuition of \S5.5.9 on incorporating the lifetime PD when rendering SICR-flagging decisions. However, this result is also unsurprising since the associated PD-model (see \cref{app:PD_Model}) has an input space that is similar (but smaller) to those of the various SICR-models. Therefore, not only do these SICR-models predict future SICR-events more accurately, but they also do so more parsimoniously than the PD-comparison approach.
Lastly, we measured the relative contribution of each input variable $x_{it}$ to the overall SICR-prediction $p_1(\boldsymbol{x}_{it})$ using \textit{Goodman-standardised coefficients} $\beta^*_{\mathrm{G}}$, as discussed by \citet{menard2004six} and \citet{menard2011standards}. These coefficients can be rank-ordered, thereby producing at least a rudimentary ordering of the relative strength (and hence importance) of each input in predicting the outcome. We found that \verb|g0_Delinq| had the largest $\beta^*_\mathrm{G}$-value (and hence greatest importance) across all SICR-models, which is sensible given its typical prominence in PD-models. Moreover, its presence within a SICR-model directly embeds the `backstop' from \S5.5.11 in IFRS 9 since an increase in the $g_0$-measure already constitutes a SICR-event. In this case, the odds ratio also increases substantially by about 900\% on average, which would correctly trigger a SICR-decision across all chosen cut-offs $c_{dsk}$.

\begin{longtable}{p{3.5cm} p{7.8cm} p{1.8cm} p{2.4cm}}
\caption{The selected input variables across the different SICR-models, given SICR-definitions from \autoref{tab:SICR_Defs}.}
\label{tab:featuresdescription} \\
\toprule
\textbf{Variable} & \textbf{Description} & \textbf{Definitions} & \textbf{Theme} \\ 
\midrule
\endfirsthead
\caption[]{(continued)} \\
\toprule
\textbf{Variable} & \textbf{Description} & \textbf{Definitions} & \textbf{Theme} \\ 
\midrule
\endhead
\midrule \multicolumn{4}{r}{\textit{Continued on next page}} \\
\endfoot
\bottomrule
\endlastfoot
\footnotesize{\verb|ArrearsDir_3|} & \footnotesize{The trending direction of the arrears balance over 3 months, obtained qualitatively by comparing the current arrears-level to that of 3 months ago, binned as: 1) increasing; 2) milling; 3) decreasing (reference); and 4) missing.} & 1a, 1b, 1c, 2a, 2b & Delinquency \\
\footnotesize{\verb|BalanceLog|} & \footnotesize{Log-transformed outstanding balance at month-end.} & 1a, 1c, 2a & Account-level \\
\footnotesize{\verb|BalanceToTerm|} & \footnotesize{Outstanding balance divided by the contractual term of the loan.} & 1b & Account-level \\
\footnotesize{\verb|DebtToIncome|} & \footnotesize{Debt-to-Income: Average household debt expressed as a percentage of household income per quarter, interpolated monthly.} & 1a, 1b, 1c, 2a, 2b & Macroeconomic \\
\footnotesize{\verb|DebtToIncome_12|} & \footnotesize{Debt-to-Income: 12-month lagged version of \verb|DebtToIncome|.} & 1a, 1b, 1c, 2a & Macroeconomic \\
\footnotesize{\verb|Employment_Growth|} & \footnotesize{Year-on-year growth rate in the 4-quarter moving average of employment per quarter, interpolated monthly.} & 2b, 2c & Macroeconomic \\
\footnotesize{\verb|g0_Delinq|} & \footnotesize{Delinquency measure: number of payments in arrears; see $g_0$-measure in \citet{botha2021paper1}.} & 1a, 1b, 1c, 2a, 2b, 2c & Delinquency \\
\footnotesize{\verb|Inflation_Growth|}  & \footnotesize{Year-on-year growth rate in inflation index (CPI) per month.} & 1a, 1b, 1c, 2a, 2b & Macroeconomic \\
\footnotesize{\verb|InterestRate_Margin|} & \footnotesize{Margin between an account's nominal interest rate and the current prime lending rate, as set by the South African Reserve Bank (SARB)} & 1a, 1b, 1c, 2a & Account-level \\
\footnotesize{\verb|PayMethod|} & \footnotesize{A categorical variable designating different payment methods: 1) debit order (reference); 2) salary; 3) payroll or cash; and 4) missing.} & 1a, 1b, 1c, 2a & Behavioural \\
\footnotesize{\verb|PD_ratio|} & \footnotesize{Ratio between two estimates of default risk at different time points: the current point in time vs initial recognition. Signifies the change in the lifetime PD; see \cref{app:PD_Model}.} & 1a, 1b, 1c, 2a, 2b, 2c & Delinquency \\
\footnotesize{\verb|PerfSpell_Num|} & \footnotesize{Current performing spell number in tracking previous default spells.} & 1a, 1b, 1c, 2a & Delinquency \\
\footnotesize{\verb|Prepaid_Pc|} & \footnotesize{The prepaid or undrawn fraction of the available credit limit.} & 1a, 1b, 1c, 2a, 2b & Behavioural \\
\footnotesize{\verb|RealGDP_Growth|} & \footnotesize{Year-on-year growth rate in the 4-quarter moving average of real GDP per quarter, interpolated monthly.} & 1a & Macroeconomic \\
\footnotesize{\verb|RealIncome_Growth|} & \footnotesize{Year-on-year growth rate in the 4-quarter moving average of real income per quarter, interpolated monthly.} & 1b, 1c & Macroeconomic \\
\footnotesize{\verb|RealIncome_Growth_12|} & \footnotesize{12-month lagged version of \verb|RealIncome_Growth|.} & 1b, 1c, 2c & Macroeconomic \\
\footnotesize{\verb|Repo_Rate|} & \footnotesize{Prevailing repurchase rate set by the South African Reserve Bank (SARB).} & 1a, 1b, 1c, 2a & Macroeconomic \\
\footnotesize{\verb|RollEver_24|} & \footnotesize{Number of times that loan delinquency increased during the last 24 months, excluding the current time point.} & 1a, 1b, 1c, 2a, 2b, 2c & Delinquency \\
\footnotesize{\verb|Term|} & \footnotesize{Contractual term of the loan.} & 1a, 1b, 1c & Account-level \\
\footnotesize{\verb|TimeInPerfSpell|} & \footnotesize{Duration (in months) of current performing spell before default or competing risk.} & 1a, 1b, 1c, 2a & Delinquency \\ 
\end{longtable}

\subsection{A basic model for generating lifetime PD-estimates}
\label{app:PD_Model}

In facilitating feature engineering and certain comparisons, we require estimates of a loan's intrinsic default risk over its lifetime. Accordingly, an elementary but typical model is built for predicting this probability of defaulting (PD) over a 12-month period, as applied at any point during loan life. 
Note that \S5.5.9 in IFRS 9 from the \citet{ifrs9_2014} technically requires the use of a lifetime PD-measure when monitoring default risk over loan life for SICR-purposes, which suggests the use of more sophisticated survival models. However, and as allayed in \S B5.5.13, a 12-month PD-measure can reasonably approximate its lifetime counterpart when the former captures the most salient of time-dependent default patterns over loan life.
Our 12-month PD-model is trained on the same data that fed the various SICR-models, and retains the same resampling scheme and sample design from \cref{app:samplingDesign}. 
Having used binary logistic regression as the underlying modelling technique (see \cref{app:logistic}), we present and describe the finalised feature space in \autoref{tab:features_PD}, which has some overlap with that of the various SICR-models.

\begin{table}[ht!]
\caption{The selected input variables of a basic PD-model; a logit-model.}
\label{tab:features_PD}
\centering
\begin{tabular}{p{3.1cm} p{6.8cm} p{3.3cm} p{2.3cm}}
\toprule
\textbf{Variable} & \textbf{Description} & \textbf{Coefficient estimate \& standard error} & \textbf{Theme} \\ 
\midrule
\footnotesize{\texttt{Intercept}} & \footnotesize{Intercept term.} & \footnotesize{-3.547479 (0.061482) } & Account-level \\
\footnotesize{\texttt{Age}} & \footnotesize{The overall loan age, measured in calendar months.} & \footnotesize{ 0.000617 (0.000107)} & Account-level \\
\footnotesize{\texttt{BalanceReal}} & \footnotesize{The inflation-adjusted outstanding balance at month-end.} & \footnotesize{0.0000015 (0.00000005)} & Account-level \\
\footnotesize{\texttt{InterestRate\_Margin}} & \footnotesize{Margin between an account's nominal interest rate and the current prime lending rate; proxy for embedding risk-based pricing principles.} & \footnotesize{9.755492 (0.656261)} & Account-level \\
\multirow{3}{3.1cm}{\footnotesize{\texttt{PayMethod}}} & \multirow{3}{6.8cm}{\footnotesize{A categorical variable designating different payment methods: 1) debit order (reference); 2) salary; 3) payroll or cash; and 4) missing.}} & \footnotesize{Missing: \par 0.831678 (0.019762)} & \multirow{3}{2.3cm}{Behavioural} \\
 & & \footnotesize{Salary: \par 0.498160 (0.027952)} & \\
 & & \footnotesize{Payroll or cash: \par 1.513412 (0.017141)} & \\
\footnotesize{\texttt{Prepaid\_Pc}} & \footnotesize{The prepaid or undrawn fraction of the available credit limit.} & \footnotesize{ -6.069346 (0.214468) } & Behavioural \\
\footnotesize{\texttt{PrincipalReal}} & \footnotesize{The inflation-adjusted loan amount originally granted.} & \footnotesize{ -0.0000013 (0.00000005)} & Account-level \\
\footnotesize{\texttt{Term}} & \footnotesize{Contractual term of the loan.} & \footnotesize{-0.000918 (0.000253)} & Account-level \\
\bottomrule
\end{tabular}
\end{table}

Central to any model is the question of its prediction accuracy and discriminatory power beyond training data. Accordingly, an ROC-analysis on the resulting probability scores of this PD-model yields an AUC-statistic of 73.94\%, having used the validation set $\mathcal{D}_V$. While certainly not stellar, the prediction accuracy is still contextually decent when considering the relatively constrained input space of this PD-model. 
The prediction accuracy can also be assessed at the portfolio-level by comparing time graphs of the actual vs expected 12-month default rates over time $t$. In particular, the expected rate is simply the mean of the model-derived default probability scores $p_1(\boldsymbol{x}_{it})$ over all performing accounts $i$ at each $t$, while the actual rate is similarly defined to the SICR-rate from \autoref{eq:SICR_Rate}. 
More formally, let $D^i_{t}$ be a period-level default indicator for account $i$ at each point $t$ of its lifetime, let $\mathcal{S}_P(t)$ denote a set of performing accounts at $t$ that are subject to default risk, and let $n'_t$ signify the number of such accounts in $\mathcal{S}_P(t)$. 
Consider then the portfolio-level 12-month conditional default probability $\mathbb{P}(D_{t+12}=1 | \ D_t=0)$ where $D_t,D_{t+1},\dots$ are Bernoulli random variables over calendar time $t$. In following the \textit{worst-ever} aggregation approach from \citet[\S 3.1.3]{botha2021phd}, we estimate a variant of this probability at a given $t$ within $\mathcal{S}_P(t)$ by using the actual default rate; itself defined as
\begin{equation} \label{eq:defaultRate_act}
    \frac{1}{n'_t}\sum_{i \ \in \ \mathcal{S}_P(t)}{ \left[ \max\left( D^i_{t}, \dots, D^i_{t+12} \right) = 1 \right]} \, ,
\end{equation}
where $[\cdot]$ are Iverson brackets. Likewise, the expected default rate is similarly defined as
\begin{equation} \label{eq:defaultRate_exp}
    \frac{1}{n'_t}\sum_{i \ \in \ \mathcal{S}_P(t)}{ p_1(\boldsymbol{x}_{it}) } \, .
\end{equation}

\begin{figure}[ht!]
\centering\includegraphics[width=0.9\linewidth,height=0.55\textheight]{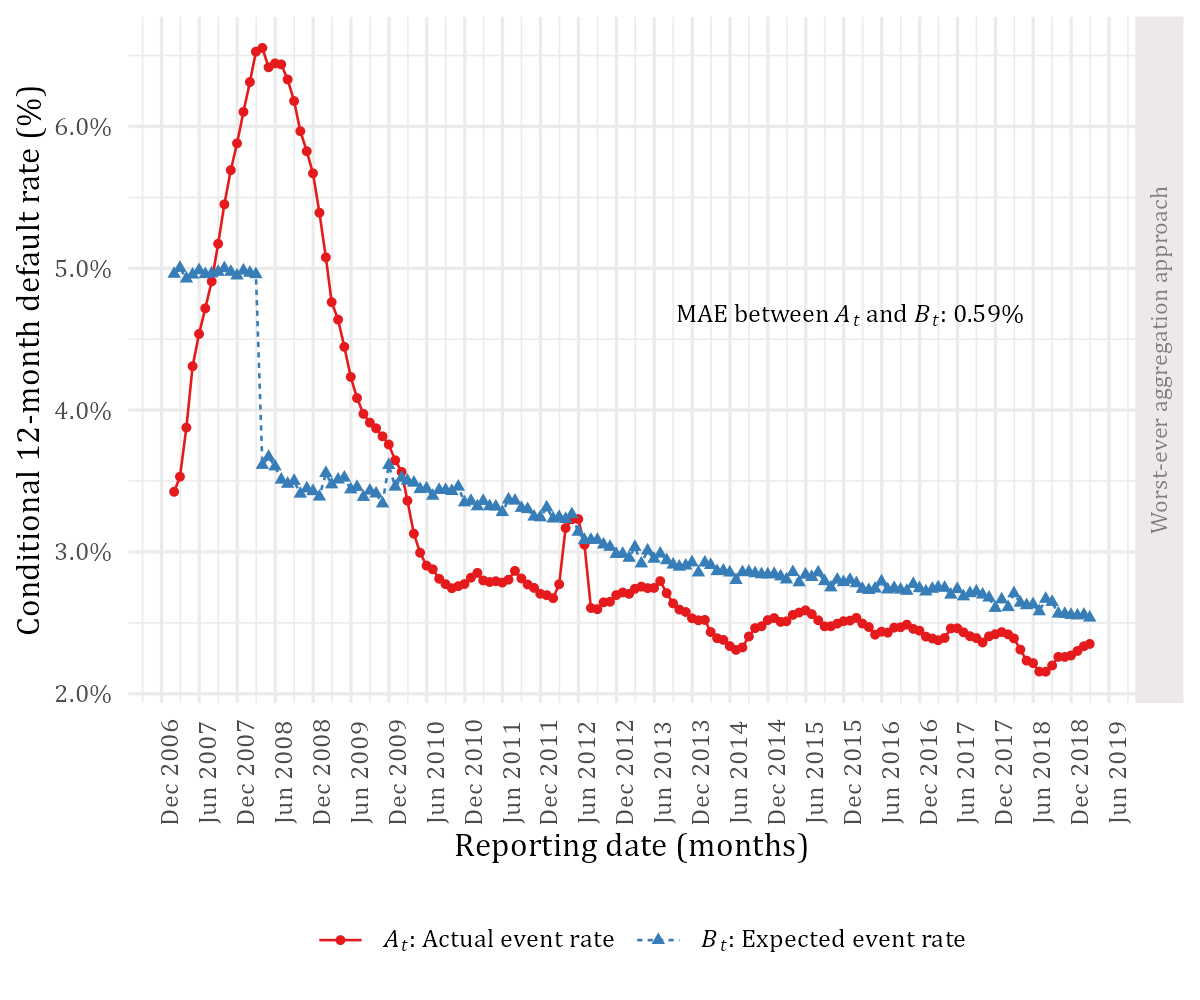}
\caption{Time graphs of various 12-month default rates within $\mathcal{D}_S$. These conditional event rates include the actual rate $A_t$, as well as the expected prediction $B_t$ thereof, as defined in \crefrange{eq:defaultRate_act}{eq:defaultRate_exp}. The Mean Absolute Error (MAE) between both rates is overlaid in summarising the discrepancies over time. }\label{fig:ActExp_Default}
\end{figure}

Using \crefrange{eq:defaultRate_act}{eq:defaultRate_exp} within the subsampled set $\mathcal{D}_S$, we present in \autoref{fig:ActExp_Default} the time graphs of the actual and expected default rates, $A_t$ and $B_t$ over time $t$. As with the SICR-rates from \autoref{sec:outcomePeriods}, both time graphs should ideally overlap with each other quite closely, thereby suggesting that the aggregated predictions agree with reality.
The early peak in $A_t$ clearly signifies the 2008-GFC, while $B_t$ reacts at least moderately in its prediction of the prevailing default rates. Aside from the anomalous 2008-GFC, the prediction $B_t$ evidently exceeds the actual default experience $A_t$ across the majority of periods, which is reassuringly risk-prudent. The average discrepancy between $A_t$ and $B_t$ is again measured using the MAE across all $t$, which yields an overall error value of 0.59\%. While the portfolio-level prediction accuracy can certainly be qualified as mediocre in the worst case, one should again consider the rather limited input space of this basic PD-model. Given its intuitive variables, its contextually decent AUC-value, and its prudential over-prediction of actual default rates, we therefore deem this PD-model as sufficient for our purposes.

\subsection{Performance measures for assessing SICR-definitions and the resulting SICR-models}
\label{app:PerfMeasures}

We formulate a few measures for evaluating various aspects of our SICR-models and their underlying SICR-definitions from \autoref{tab:SICR_Defs}, as characterised by the triple $(d,s,k)$. Some of these measures focus on SICR-predictions at the loan account-level (as summarised over time), while others are defined at the portfolio-level; both aggregation levels provide useful perspectives.
These performance measures can also be applied more generally on the predictions and/or decisions of any binary SICR-classification system under IFRS 9. Doing so can foster comparability with our results, as well promote standardisation across the industry when evaluating or auditing SICR-decisions.

\textbf{SICR-prevalence:} Denoted as $\phi_{dsk}$, the prevalence estimates the prior class probability $\mathbb{P}\left(Y=1\right)$ for a Bernoulli random variable $Y$ that represents the portfolio-level SICR-outcome, as given by the $\mathcal{Z}_t(d,s,k)$-process from \autoref{eq:decision_rule_generator}. 
Assume a sample $\mathcal{D}=\left\{i,t,y_{it}\right\}$ of binary-valued SICR-outcomes $y_{it}$ that are observed for accounts $i=1,\dots,N$ at each period $t=t_1,\dots,T_i-k$ over each account's lifetime $T_i$ from its time of initial recognition $t_1$. Given $\mathcal{D}$ of size $n=|\mathcal{D}|$, we estimate the prevalence using Iverson brackets $[\cdot]$ as
\begin{equation} \label{eq:SICR_Prevalence}
    \phi_{dsk} = \frac{1}{n} \sum_{it \, \in \, \mathcal{D}}{\left[ y_{it} =1 \right]} \, .
\end{equation}
Put differently, $\phi_{dsk}$ is the proportion of rare events in $\mathcal{D}$, which also measures the degree of class imbalance. A SICR-model is then eventually built using these $y_{it}$-values as the target/outcome variable.

\textbf{SICR-rate:} Denoted as $A_t$, the SICR-rate estimates at reporting/calendar time $t$ the portfolio-level transition probability of moving from Stage 1 to Stage 2 impairment over a $k$-month period. More formally, $A_t$ estimates at $t$ the conditional probability $\mathbb{P}\left(Y_{t+k}= 1 | \, Y_t = 0 \right)$ of becoming SICR-flagged later at $t+k$, where $Y_t,Y_{t+1},\dots$ are Bernoulli random variables that represent the SICR-status over $t$, as given by $\mathcal{G}(d,s,t)$ from \autoref{eq:bool_decision}. 
Given the previous sample $\mathcal{D}$ for estimating the prevalence $\phi_{dsk}$-measure, we can partition $\mathcal{D}$ into non-overlapping subsets $\mathcal{S}_1(t)$ over reporting time $t=1,\dots$, where each $\mathcal{S}_1(t)\in\mathcal{D}$ contains all Stage 1 accounts $i$ at $t$ that are at risk of becoming SICR-flagged later at $t+k$. More formally, each $\mathcal{S}_1(t)$-subset is defined at a given reporting time $t$ as 
\begin{equation} \label{eq:SICR_AtRiskSet}
    \mathcal{S}_1(t) = \left\{i : \mathcal{G}_i(d,s,t) = 0 \right\} \quad \text{for } i \in\mathcal{D} \, ,
\end{equation}
where $\mathcal{G}_i(d,s,t)$ simply denotes the SICR-status at the corresponding loan period $t$ of account $i$, as calculated using \autoref{eq:bool_decision}. Given a risk set $\mathcal{S}_1(t)$ of size $n_t=|\mathcal{S}_1(t)|$, we then estimate the SICR-rate $A_t$ at a given reporting period $t$ by using Iverson brackets $[\cdot]$ on the binary-valued SICR-outcomes $y_{it}\in\{0,1\}$ within $\mathcal{S}_1(t)$, expressed  as 
\begin{equation} \label{eq:SICR_Rate}
    A_t = \frac{1}{n_t}\sum_{it \, \in \, \mathcal{S}_1(t)}{ \left[y_{it} = 1 \right]} \, .
\end{equation}

While \autoref{eq:SICR_Rate} yields the \textit{actual} SICR-rate over time, the account-level predictions from an underlying SICR-model can be similarly aggregated into an \textit{expected} SICR-rate $B_t$, which can be duly compared to $A_t$. This $B_t$-quantity similarly estimates at $t$ the conditional probability $\mathbb{P}\left(Y_{t+k}= 1 | \, Y_t = 0, \boldsymbol{X} \right)$ of becoming SICR-flagged later at $t+k$, given the random input vector $\boldsymbol{X}$. Estimating this probability implies developing a SICR-model from data, which can then be used to render predictions on new data.
In particular, a SICR-prediction refers to the model-derived probability score $p_1(\boldsymbol{x}_{it})\in[0,1]$ in predicting the SICR-outcome $y_{it}$ using multivariate input data $\boldsymbol{x}_{it}$.
Assume a scored sample $\mathcal{D}=\left\{i,t,y_{it}, p_1(\boldsymbol{x}_{it})\right\}$ of such $p_1(\boldsymbol{x}_{it})$-scores that are estimated given data $\boldsymbol{x}_{it}$ of accounts $i=1,\dots,N$ at each period $t=t_1,\dots,T_i-k$.
This $\mathcal{D}$ is again partitioned into subsets $\mathcal{S}_1(t)$ over $t$ using \autoref{eq:SICR_AtRiskSet}, whereupon the expected SICR-rate $B_t$ is estimated at $t$ similarly to \autoref{eq:SICR_Rate} as the mean score, defined as
\begin{equation} \label{eq:SICR_Rate_exp}
    B_t = \frac{1}{n_t}\sum_{it \, \in \, \mathcal{S}_1(t)}{ p_1(\boldsymbol{x}_{it})} \, .    
\end{equation}
We formulate another variety of \autoref{eq:SICR_Rate_exp} called the \textit{discretised} expected SICR-rate $C_t$, wherein the underlying SICR-model (itself a discriminative/probabilistic classifier) is first dichotomised into a discrete classifier. This dichotomisation requires evaluating each SICR-prediction against a static cut-off $c_{dsk}\in[0,1]$, thereby producing a discrete SICR-prediction. Given the same risk set $\mathcal{S}_1(t)$, we estimate the discretised expected SICR-rate $C_t$ at a given reporting time $t$ and using Iverson brackets $[\cdot]$ as 
\begin{equation} \label{eq:SICR_Rate_expDisc}
    C_t = \frac{1}{n_t}\sum_{it \, \in \, \mathcal{S}_1(t)}{ \left[ p_1(\boldsymbol{x}_{it}) > c_{dsk} \right]} \, .     
\end{equation}

\textbf{SICR-mean:} This quantity is simply the sample mean of portfolio-level SICR-rates over time. Given the actual SICR-rates $A_t$ from \autoref{eq:SICR_Rate} over reporting time $t=1,\dots,t_n$, we estimate the SICR-mean as $\bar{A} = (t_n)^{-1} \sum_{t}{A_t}$.

\textbf{Instability:} Denoted as $\sigma_{dsk}$, the instability refers to the degree to which a series of portfolio-level SICR-rates varies over time. Given the actual SICR-rates $A_t$ from \autoref{eq:SICR_Rate} over reporting time $t=1,\dots,t_n$, we estimate the instability $\sigma_{dsk}$ using the SICR-mean $\bar{A}$ and the sample standard deviation as
\begin{equation} \label{eq:SICR_Rate_stdev}
    \sigma_{dsk} = \sqrt{\frac{1}{t_n-1}\sum_{t=1}^{t_n} {\left( A_t - \bar{A} \right)^2} } \, .
\end{equation}

\textbf{Prediction dynamicity:} Denoted as $\omega_{dsk}$, the dynamicity represents the extent to which the SICR-predictions (or probability scores) vary over the lifetime of an average loan account.
Assume a sample $\mathcal{D}=\left\{i,t, p_1(\boldsymbol{x}_{it})\right\}$ of model-derived probability scores $p_1(\boldsymbol{x}_{it})$ that are estimated for accounts $i=1,\dots,N$ at each period $t=t_1,\dots,T_i-k$ over each account's lifetime $T_i$ from its time of initial recognition $t_1$, given data $\boldsymbol{x}_{it}$.
Having calculated the standard deviation $\omega_i$ of all the $p_1(\boldsymbol{x}_{it})$-scores over the lifetime of each account $i$, this $\omega_{dsk}$-quantity is then estimated by taking the sample mean of these $\omega_i$-estimates. More formally, we first estimate each $\omega_i$ by calculating the account-level standard deviation of scores for each account $i$ in a given sample $\mathcal{D}$, expressed as
\begin{equation} \label{eq:Prediction_stdev}
    \omega_i = \sqrt{\frac{1}{n_i-1}\sum_{it \, \in \, \mathcal{D}} {\left( p_1(\boldsymbol{x}_{it}) - \left\{\frac{1}{n_i} \sum_{it \, \in \, \mathcal{D}}{p_1(\boldsymbol{x}_{it})} \right\} \right)^2} } \, ,
\end{equation}
where $n_i=T_i-k>0$ denotes the number of probability scores that are available for account $i$ in $\mathcal{D}$.
Given this new account-level sample $\mathcal{D}_N=\{i,\omega_i\}$ of standard score deviations $\omega_1,\dots,\omega_N$, we finally estimate $\omega_{dsk}$ as
\begin{equation} \label{eq:Prediction_dynamicity}
    \omega_{dsk} = \frac{1}{N} \sum_{i \, \in \mathcal{D}_N}{\omega_i} \, .
\end{equation}


\singlespacing
\printbibliography 

\onehalfspacing



\end{document}